\documentclass[twocolumn,aps,prd,showpacs,nofootinbib,superscriptaddress,preprintnumbers,floatfix,10pt]{revtex4-1}

\usepackage{amsmath}
\usepackage{amsfonts}
\usepackage{amssymb}
\usepackage{epsfig}
\usepackage{graphicx}
\usepackage{color}
\usepackage{slashed}
\usepackage{epstopdf}

\newcommand{\STr}{\text{STr}}

\newcommand{\fss}[1]{#1\!\!\!/}   
\newcommand{\Eqref}[1]{Eq.~\eqref{#1}}
\newcommand{\Nc}{N_{\mathrm{c}}}
\newcommand{\Nf}{N_{\mathrm{f}}}
\newcommand{\pt}{\partial_t}
\newcommand{\yb}{\bar{\psi}}

\newcommand{\lL}{\lambda_\Lambda}
\newcommand{\mH}{m_{\text{H}}}
\newcommand{\mtop}{m_{\text{t}}}
\newcommand{\mbot}{m_{\text{b}}}
\newcommand{\htop}{h_{\text{t}}}
\newcommand{\hbot}{h_{\text{b}}}
\newcommand{\Np}{N_{\text{p}}}

\begin{document}

\preprint{}

\title {
Higgs Mass Bounds from Renormalization Flow for a Higgs-top-bottom model} 

\author{Holger Gies}
\email{holger.gies@uni-jena.de}
\affiliation{Theoretisch-Physikalisches Institut, Friedrich-Schiller-Universit\"at Jena, Max-Wien-Platz 1, 
D-07743 Jena, Germany}
\author{Ren\'{e} Sondenheimer}
\email{rene.sondenheimer@uni-jena.de}
\affiliation{Theoretisch-Physikalisches Institut, Friedrich-Schiller-Universit\"at Jena, Max-Wien-Platz 1, 
D-07743 Jena, Germany}

\begin{abstract}
  We study a chiral Yukawa model mimicking the Higgs-top-bottom sector
  of the standard model. We re-analyze the conventional arguments that
  relate a lower bound for the Higgs mass with vacuum stability in the
  light of exact results for the regularized fermion determinant as
  well as in the framework of the functional renormalization group. In
  both cases, we find no indication for vacuum instability nor
  meta-stability induced by top-fluctuations if the cutoff is kept
  finite but arbitrary. A lower bound for the Higgs mass arises for
  the class of standard bare potentials of $\phi^4$ type from the
  requirement of a well-defined functional integral (i.e., stability
  of the bare potential). This consistency bound can however be 
  relaxed considerably by more general forms of the bare potential
  without necessarily introducing new meta-stable minima.
\end{abstract}


\maketitle

\section{\label{intro} Introduction}

Long before the recent discovery of a comparatively light
standard-model Higgs boson \cite{Aad:2012tfa}, estimates and bounds on
this mass parameter have been derived from renormalization arguments
\cite{Maiani:1977cg,Krasnikov:1978pu,Lindner:1985uk,Wetterich:1987az,Altarelli:1994rb,Schrempp:1996fb,Hambye:1996wb}. Assuming
the validity of the standard model over a wide range of scales up to
an ultraviolet (UV) cutoff scale $\Lambda$, together with mild
assumptions on the microscopic action at the scale $\Lambda$,
typically leads to a finite range of possible low-energy values for
the Higgs mass, the so-called infrared (IR) window
\cite{Wetterich:1987az,Wetterich:1981ir}. Similar arguments can also
be applied to models beyond the standard model of particle physics
\cite{Cabibbo:1979ay,Espinosa:1991gr,Kadastik:2011aa,Chen:2012faa,Lebedev:2012zw}. It
has been suggested that Higgs masses below the lower bound necessarily
require the effective potential of the standard model to develop a
further minimum beyond the electroweak minimum
\cite{Krive:1976sg,Hung:1979dn,Linde:1979ny,Politzer:1978ic,Sher:1988mj,Lindner:1988ww,Ford:1992mv}. Since
the measured value of the Higgs mass near $\mH=125$GeV appears to be
near if not below the lower bound, the standard model vacuum could be
unstable or at least metastable. In the latter case, the metastability
has to be sufficiently long-lived compared to the age of the universe
to allow for our existence
\cite{Arnold:1989cb,Sher:1993mf,Bergerhoff:1999jj,Isidori:2001bm,Ellis:2009tp,EliasMiro:2011aa,Degrassi:2012ry,Alekhin:2012py,Masina:2012tz,Buttazzo:2013uya}.

While the occurrence of the vacuum instability is often attributed to
the fluctuation of the top quark (and therefore sensitively depends
on the top mass), the conventional perturbative analysis of
determining the instability has been questioned by non-perturbative
methods. Within the toy model of a top-Higgs-Yukawa system with
discrete symmetry, lattice simulations have revealed that the full
effective potential in this model with the cutoff kept finite does not
develop an instability \cite{Holland:2003jr,Holland:2004sd,Fodor:2007fn}. By
contrast, the perturbative treatment of the same model in the limit
$\Lambda\to\infty$ exhibits an instability in disagreement with the
simulation results. Within the same model and using functional
methods, the occurrence of the erroneous instability has been traced
back to an implicit renormalization condition that contradicts the
underlying assumption of a well-defined functional integral
\cite{Branchina:2005tu,Gneiting:2005}.

In a series of recent lattice simulations using chiral Higgs-Yukawa
models, imposing the criterion of a stable bare potential (typically
of $\phi^4$ type) has lead to a number of quantitative results for the
lower bound on the Higgs mass \cite{Fodor:2007fn,Gerhold:2007yb}
without the need to require low-energy stability. The same line of
argument can in fact be used to put strong constraints on the
existence of a fourth generation of flavor in the light of the Higgs
boson mass measurement \cite{Gerhold:2010wv,Bulava:2012pb}. These
results have also been substantiated by conventional analytical
methods \cite{Djouadi:2012ae}.

In a recent work, we have been able to show that the sole
consideration of bare potentials of $\phi^4$ type is actually too
restrictive \cite{Gies:2013fua}. In fact, if the standard model is
viewed as a low-energy effective theory, there is no reason to exclude
higher-dimensional operators from the bare potential. Their occurrence
is actually expected. Whereas Wilsonian renormalization group (RG) arguments of course suggest
that low-energy observables remain almost completely unaffected by the
higher-dimensional operators, we have demonstrated that Higgs-mass
bounds can in fact exhibit a significant dependence on the bare
potential. This may seem counter-intuitive at first sight, since the
Higgs mass is clearly an IR observable. However, a Higgs-mass bound
formulated in terms of a function of the cutoff,
$\mH{}_{,\text{bound}}(\Lambda)$, can be strongly influence by a nontrivial
RG running of the couplings near the cutoff. In \cite{Gies:2013fua},
we have identified a simple RG mechanism that leads to a lowering of
the conventional lower bound for the case of a
$\mathbb{Z}_2$-symmetric Yukawa toy model. 

Our findings of a ``lowering'' of the lower bound have been confirmed
in the chiral Yukawa models studied on the lattice
\cite{Hegde:2013mks} as well as in a model involving an additional
dark-matter scalar \cite{Eichhorn:2014qka}.

In view of the latent controversy about the (in-)existence of a
top-fluctuation induced vacuum in-/meta- stability of the electroweak
vacuum, the purpose of this work is twofold: first, we demonstrate on
the basis of exact results for the fermion determinant in the presence
of a scalar vacuum expectation value that the interaction part
contribution to the effective potential is positive if the UV cutoff
is kept finite but arbitrary. By contrast, a removal of the cutoff by
taking a naive limit $\Lambda\to\infty$ makes the fermionic
contribution to the effective potential unstable.  Second, we
generalize our results of \cite{Gies:2013fua} to the chiral Yukawa
model that is also used in lattice simulations. Using the functional
RG, we demonstrate that the conventional lower-bound can be relaxed
considerably by more general forms of the bare potential without
necessarily introducing new meta-stable minima.

The paper is organized as follows: in Sect. \ref{sec:model}, we
introduce our model and all relevant notation. Section
\ref{sec:determinant} is devoted to an analysis of exact properties of
the fermion determinant in order to explore the origin of the
apparent instability of the standard model vacuum. The
nonperturbative RG flow equations of the present model are summarized
in Sect.~\ref{sec:flow}. These are used in Sect.~\ref{sec:mHbounds} to
compute the Higgs mass bounds of the model nonperturbatively for
various bare potentials. Conclusions are presented in
Sect.~\ref{sec:conc}.

\section{Chiral Higgs-top-bottom model}
\label{sec:model}
In order to illustrate our main points in a transparent fashion, we
investigate a chiral Higgs-Yukawa model forming a self-contained
subset of the standard-model. The field content consists of a scalar
field which is a complex SU$(2)$-doublet
\begin{align}
 \phi = \frac{1}{\sqrt{2}} \begin{pmatrix} \phi_1+i\phi_2\\ \phi_4+i\phi_3 \end{pmatrix},
 \label{scalarfield}
\end{align}
and two Dirac fermions which represent the top and bottom quark. The
left-handed components of the bottom and top transform as a doublet
under SU(2) while the right-handed components are singlets.
\begin{align*}
 \psi_{\text{L}} = \begin{pmatrix} t_{\text{L}}\\ b_{\text{L}} \end{pmatrix}, \quad t_{\text{R}}, \quad b_{\text{R}}.
\end{align*}

The classical euclidean action of the model is given by
\begin{equation}
\begin{split}
 S &= \! \int d^4x \Big[ \partial_{\mu}\phi^{\dagger}\partial^{\mu}\phi + U(\bar\rho) \\
&\qquad + i\bar{\psi}_{\text{L}}\slashed{\partial}\psi_{\text{L}} + i\bar{t}_{\text{R}}\slashed{\partial}t_{\text{R}} + i\bar{b}_{\text{R}}\slashed{\partial}b_{\text{R}} \\
&\qquad + i\bar{h}_{\text{b}}(\bar{\psi}_{\text{L}}\phi b_{\text{R}} + \bar{b}_{\text{R}}\phi^{\dagger}\psi_{\text{L}}) \\
&\qquad + i\bar{h}_{\text{t}}(\bar{\psi}_{\text{L}}\phi_{\mathcal{C}} t_{\text{R}} + \bar{t}_{\text{R}}\phi_{\mathcal{C}}^{\dagger}\psi_{\text{L}})
\Big].
\label{classicalaction}
\end{split}
\end{equation}
where $\phi_{\mathcal{C}}=i\sigma_2\phi^*$ denotes the charge
conjugated scalar. The scalar field couples to the fermions
via a chiral Yukawa interaction where $\bar{h}_{\text{t}}$ and $\bar{h}_{\text{b}}$
are the (bare) Yukawa couplings for the top and bottom
respectively. Furthermore, we include scalar self-interactions encoded
in the scalar potential which depends on the field invariant
\begin{equation}
\bar\rho= \phi^{\dagger}\phi.
\end{equation}

The action \eqref{classicalaction} is invariant under the following
global symmetry transformations
\begin{align*}
 \phi \rightarrow e^{i\alpha^i\frac{\sigma^i}{2}}\phi, \quad
 \psi_{\text{L}} \rightarrow e^{i\alpha^i\frac{\sigma^i}{2}}\psi_{\text{L}}, \quad
 t_{\text{R}} \rightarrow t_{\text{R}}, \quad b_{\text{R}} \rightarrow b_{\text{R}},
\end{align*}
where $\sigma^i$ are the Pauli matrices acting on the SU(2) doublet structure and
\begin{align*}
\begin{aligned}
 \phi &\rightarrow e^{i\beta_{\text{s}}}\phi, \\
 \psi_{\text{L}} &\rightarrow e^{i\beta_{\text{L}}}\psi_{\text{L}},
\end{aligned}
\qquad
\begin{aligned}
 t_{\text{R}} &\rightarrow e^{i\beta_{\text{R}}^\text{t}}t_{\text{R}}, \\
 b_{\text{R}} &\rightarrow e^{i\beta_{\text{R}}^\text{b}}b_{\text{R}}.
\end{aligned}
\end{align*}
Here, the $\beta$ angles are related to a single angle by the usual
hypercharge assignments, $\beta_{\text{s}}=\frac{1}{2} \beta$,
$\beta_{\text{L}}=\frac{1}{6}\beta$, $\beta_{\text{R}}^{\text{t}} =
\frac{2}{3} \beta$, and $\beta_{\text{R}}^{\text{b}} = -\frac{1}{3}
\beta$. Thus, the model has a global SU(2)$\times$U(1) symmetry. The
symmetry can be spontaneously broken down to a global U(1) by a
nonzero vacuum expectation value of the scalar field $\phi\to v$,
giving rise to Dirac masses for the fermions and the Higgs boson mass.

The classical action is already equipped with a potential for the
scalar field $U(\bar{\rho})$. Symmetry breaking in the quantum theory
occurs, if the corresponding renormalized potential $U$ develops a nonvanishing
minimum $\bar{\rho}_0$. In this case, we can write the
masses \footnote{For the derivative expansion used below, these mass
  definitions already agree with the pole masses.} in terms of this
minimum and the renormalized Yukawa couplings $\htop$ and $\hbot$,
\begin{align*}
 \begin{aligned}
 v &= \sqrt{2}Z_{\phi}^{\frac{1}{2}}\left<\phi\right> = \sqrt{Z_{\phi}\bar{\rho}_0}, \\
 \mH^2 &= v^2\frac{U''(\bar{\rho}_0)}{Z_{\phi}^2},
 \end{aligned}
 \quad
 \begin{aligned}
 \mtop^2 &= \frac{1}{2}v^2\htop^2, \\ \mbot^2 &= \frac{1}{2}v^2\hbot^2,
 \end{aligned}
\end{align*} 
where we already accounted for a wave function renormalization
$Z_{\phi}$ to be defined together with the other renormalized quantities below.

Apart from the missing further matter and flavor content, our model
also ignores the gauge sectors of the standard model. This avoids not
only technical complications and subtle issues arising from the
gauge-Higgs interplay \cite{Frohlich:1981yi,Maas:2012tj} in the
standard model. But it will, of course, also lead to decisive
differences to standard model properties, which will be commented on
in the course of this work. Nevertheless, the gauge sector is less
important for the main points of the present work. In order to make
closer contact with the standard model language, we fix $v=246$GeV,
$\mtop=173$GeV and $\mbot=4.2$GeV for illustrative purposes, but
leave the Higgs mass as a free parameter for the moment.

For the following discussion it is important to note that the standard
model in its conventional form (as well as the present model) may not
be extendible to arbitrarily high momentum scales. The problem of
triviality \cite{Wilson:1973jj} --  where substantial evidence has been
accumulated for $\phi^4$-type theories -- is likely to extend to the full
chiral-Yukawa sector as well. If so, the definition of the model
unavoidably requires a UV cutoff $\Lambda$ which physically plays the
role of the scale of maximum UV extent up to which a quantum field
theory description is appropriate. If the cutoff scale is sufficiently
large, Wilsonian renormalization guarantees that the IR physics
essentially depends only on a finite number of relevant and marginal
parameters, rendering the theory predictive (in spite of our ignorance
about the physics beyond $\Lambda$).

In fact, the strategy of \textit{perturbative} renormalization manifestly
allows to implicitly or explicitly take the limit $\Lambda\to\infty$
for certain physical observables. For the general definition of the
theory, it is however important to accept the fact that the cutoff
$\Lambda$ may unavoidably have to be kept finite.\footnote{In
  principle, perturbative predictions for $\Lambda\to\infty$ may
  differ from those with a finite cutoff. However, let
  $m_{\text{Obs}}$ denote the scale of a typical IR observable; then
  this difference is typically of order
  $(m_{\text{Obs}}/\Lambda)^{(2p)}$, where $p$ is some integer. For
  sufficiently large $\Lambda$, this difference hence becomes insignificant.}

Fixing physical parameters such as $v,\mtop$, and $\mbot$ is
technically implemented by renormalization
conditions. Phenomenologically, it is useful to fix these conditions
at observational IR scales. Conceptually however, it is equally well
possible to impose suitable renormalization conditions at the
UV cutoff $\Lambda$. For instance for the present model, we can fix
$\htop$, $\hbot$, and the scalar potential $U$ in terms of their bare
quantities at the cutoff, $\htop{}_{\Lambda}=\bar{h}_{\text{t}}$,
$\hbot{}_{\Lambda}=\bar{h}_{\text{b}}$, and $U_{\Lambda}$. In
practice, the fixing can be done such that the constraints set by the
physical values of $v,\mtop$, and $\mbot$ are satisfied.

From this viewpoint, the Higgs boson mass as the remaining free
parameter becomes a function of the unconstrained combinations of the
UV couplings. Now, bounds on the Higgs boson mass arise, if all
permissible choices of UV couplings result in a finite range of Higgs
boson masses. Examples for such finite ranges of possible Higgs
masses, so-called IR windows, are shown below in
Figs. \ref{fig:IRWindowHiggsMassWithinPhi4} and
\ref{fig:HiggsMassBoundsWithinPhi4}. Before we derive these results 
from the functional RG, let us specifically pay attention to the lower
bound that has been associated with vacuum stability.

\section{Fermion determinant and (in-)stability}
\label{sec:determinant}

Current state-of-the-art is a next-to-next-to-leading order analysis
of the standard model effective potential, in order to study it's
stability properties, see, e.g.,
\cite{Degrassi:2012ry,Zoller:2014xoa}. 
However, the stability issue is already visible at one-loop order. In
a nutshell, the argument goes as follows, see, e.g.,
\cite{Espinosa:2013lma}. First, standard renormalization conditions on
the effective potential are imposed in such a way that a first
nontrivial minimum occurs at $v\simeq246$GeV. This involves in
particular to choose the counterterm $\sim \phi^\dagger \phi$
appropriately. Then, one assumes based on RG-improvement arguments
that the effective potential for large field values $\phi\gg v$ is
well approximated by\footnote{In this section, we do not distinguish
  between bare and renormalized scalar fields, as this is not relevant
  at this order.}
\begin{equation}
U_{\text{eff}}(\rho)\simeq \frac{\lambda(\mu)}{2} \rho^2, \quad \rho = \phi^\dagger \phi, 
\label{eq:Ueff}
\end{equation}
where the RG scale $\mu$ then should be identified with the field
amplitude $\mu\simeq |\phi|$. The dependence of the $\phi^4$ coupling on
the scale $\mu$ can be computed by integrating its $\beta$ function
from the IR (where the boundary condition is fixed in terms of the
Higgs mass) upwards to higher scales. For the present model, the $\beta$ function reads,
\begin{equation}
\beta_\lambda =  \frac{1}{4 \pi^2} \left[ - \htop^4 - \hbot^4 +
   \lambda (\htop^2 +\hbot^2) + 3\lambda^2 \right]. \label{eq:betapert}
\end{equation}
(This agrees with the corresponding standard-model sectors if a
  color factor of $\Nc$ is included for each fermion loop.)
Obviously, the pure fermion-loop contributions $\sim\htop^4,\hbot^4$
come with a negative sign, implying that they tend to deplete
$\lambda$ towards the UV. In particular for a heavy top (large
$\htop$) and a light Higgs (small $\lambda$), the integrated coupling
$\lambda(\mu)$ can drop below zero at high scales. Identifying $\mu$
with $|\phi|$ and inserting this result into \Eqref{eq:Ueff} leads us
to the standard conclusion that the effective potential seems to
develop an instability towards large field values. [NB: In the present
  model, this line of argument actually results in an effective
  potential being unbounded from below for $|\phi| \to \infty$. In the
  full standard model, electroweak fluctuations eventually stabilize
  this effective potential again by turning $\lambda(\mu)$ back to
  positive values for scales typically far above the Planck scale, see,
  e.g. \cite{Gabrielli:2013hma}.]

With hindsight, the arguments underlying \Eqref{eq:Ueff} rely on the
assumption that the field amplitude $\phi$ provides for the only
relevant scale at large values. It is precisely this assumption that
fails in the presence of a finite cutoff independently of the size of
the cutoff. In order to see this, let us start from the (for
simplicity Euclidean) generating functional for the scalar correlation functions of the model
\begin{equation}
Z[J]= \int_\Lambda \mathcal{D}\phi \mathcal{D} \bar\psi \mathcal{D} \psi e^{-S[\phi,\yb,\psi]+\int J \phi},
\label{eq:genfunc}
\end{equation}
where the appearance of $\Lambda$ at the functional integral shall
remind us of the fact that the theory requires a regularization
procedure as part of its definition. As the action is a quadratic form
in the fermion fields, the corresponding fermionic integral can be
carried out and yields
\begin{eqnarray}
Z[J]&=& \int_\Lambda \mathcal{D}\phi\, \text{det}_\Lambda (\mathcal{L}(\phi))  e^{-S_{\text{B}}[\phi]+\int J \phi} \nonumber\\
&=&\int_\Lambda \mathcal{D}\phi\, e^{-S_{\text{B}}[\phi]-S_{\text{F},\Lambda}[\phi] +\int J\phi},
\label{eq:genfunc2}
\end{eqnarray}
where $S_{\text{B}}$ is the purely bosonic part of the action and
$\mathcal{L}(\phi)$ denotes the Dirac operator in the presence of the
scalar field. In the second line of \Eqref{eq:genfunc2}, we have
introduced the effective action $S_{\text{F},\Lambda}[\phi]$ arising
from integrating out the fermion fluctuations. As the fermion
determinant and thus also $S_{\text{F},\Lambda}$ corresponds already
to a loop-integration, it suffices for the present purpose to
investigate its properties for a homogeneous \textit{mean field}
$\phi$. Deviations from this mean field contribute to the full
effective potential only in terms of fluctuations at higher-loop
order. Therefore, we concentrate on the fermion-fluctuation induced
contribution to the effective potential
\begin{equation}
U_{\text{F}}(\rho) = - \frac{1}{\Omega} \ln \text{det}_{\Lambda} (\mathcal{L}(\phi)),
\label{eq:UF}
\end{equation}
where $\Omega=\int d^4x$ denotes the spacetime volume, and we have used the
fact that the dependence on $\phi$ must occur in terms of the SU(2)
invariant variable $\rho$. Incidentally, \Eqref{eq:UF} corresponds to
the leading contribution to the effective potential at large $\Nf$.

Upon a global SU(2) rotation, the mean field can be rotated into the
$\phi_4$ component of the scalar field. The Dirac operator then
becomes block diagonal in top-bottom space, reading
$\mathcal{L}_a(\phi) = i \slashed{\partial} + i \frac{1}{\sqrt{2}} h_a
\phi_4$ where $a=\{\text{t},\text{b}\}$ in the top or bottom subspace
respectively. Because of $\gamma_5$ hermiticity, $i\slashed{\partial}$
is isospectral to $-i\slashed{\partial}$ which allows us to write
\begin{equation}
U_{\text{F}}(\rho) = - \frac{1}{2\Omega}\sum_{a=\{\text{t},\text{b}\}}
\ln \frac{\text{det}_{\Lambda} ( -\partial^2 +h_a^2 \rho)}
    {\text{det}_{\Lambda} ( -\partial^2)},
\label{eq:UF2}
\end{equation}
where $\rho=\frac{1}{2} \phi_4^2$ for our choice of mean field. In
proceeding from \Eqref{eq:UF} to \Eqref{eq:UF2}, we also used the
freedom of choosing the normalization of the generating functional
such that the fermion-induced effective potential is normalized to the
zero-field limit, i.e., $U_{\text{F}}(\rho=0)=0$. This resulting ratio
of determinants can be evaluated straightforwardly, once a
regularization procedure has been chosen. The final result will of
course depend on the regularization for any finite value of the cutoff
$\Lambda$. As argued above, we should not expect that the cutoff can
be sent to infinity, since our model is likely to have a scale of
maximum UV extent. In order to understand this regulator dependence,
it is therefore instructive to compute \Eqref{eq:UF2} for different
choices of the regularization.

\subsubsection{Sharp cutoff}

As the contributions from both quarks are formally identical
up to a different value of the Yukawa coupling, it is sufficient to
study $U_{\text{F},a}$ and take the sum over
${a=\{\text{t},\text{b}\}}$ afterward. A straightforward
regularization is provided by a sharp cutoff in momentum space, such
that \Eqref{eq:UF2} translates into
\begin{equation}
U_{\text{F},a}(\rho) = - 2 \int_\Lambda \frac{d^4p}{(2\pi)^4}
\ln \left( 1+ \frac{h_a^2\rho}{p^2} \right) ,
\label{eq:SC1}
\end{equation}
where we have used that we work here with 4-component Dirac
spinors. As expected, the integral contains quadratic and logarithmic
``divergencies'', which can be made explicit by writing the analytic
exact result of the integral as
\begin{eqnarray}
U_{\text{F},a}(\rho) &=&
-\frac{\Lambda^2}{8\pi^2} h_a^2 \rho \label{eq:SC1a} \\
&&+ \frac{1}{16\pi^2} \left[h_a^4\rho^2 \ln\left(\! 1+ \frac{\Lambda^2}{h_a^2\rho}\! \right)
+h_a^2\rho \Lambda^2 \right.
\nonumber\\
&&\left. \qquad\qquad- \Lambda^4 \ln\left(\! 1+ \frac{h_a^2\rho}{\Lambda^2}\! \right)\right].\label{eq:SC1b}
\end{eqnarray}
Here observe that the quadratic divergence $\sim\Lambda^2$ has been
isolated in the first line. The remaining term in square brackets
contains only logarithmic divergencies $\sim \ln\Lambda$.  It is
however more important to note that the first line also isolates the
only term proportional to $\rho\sim \phi^\dagger\phi$ and thus
contributes to the mass parameter of the scalar field. The remaining
terms represent the interacting part of the fermion-induced effective
potential.

Most importantly: whereas the contribution to the mass term is
negative, as it should be, since fermion fluctuations tend to induce
chiral symmetry breaking, the whole interaction part in square
brackets is strictly positive for all $\rho>0$. This follows
immediately from the inequality $\ln (1+x)<x$ (for $x>0$) applied to the
last term. Similarly, it can be shown that also the derivative of the
interacting part with respect to $\rho$ is strictly positive for any
finite value of $\rho$, $h_a$ and $\Lambda$.

We conclude that the fermion determinant -- apart from its
contribution to the scalar mass term -- is strictly positive and
monotonically increasing in its interacting part. Therefore, once the
scalar mass term has been fixed by a renormalization condition, the
remaining contributions from the top fluctuations to the interacting
part of the bosonic potential are strictly positive. This excludes the
possibility that an instability beyond the electroweak vacuum is
induced by fermionic fluctuations. This can also be phrased in terms
of a more rigorous statement: if the potential of the purely bosonic
part $S_\text{B}$ of the action in \Eqref{eq:genfunc2} is bounded from
below by a function of the form $U_{\text{B}}(\rho)> c_1 + c_2
\rho^{1+\epsilon}$ with an arbitrary finite constant $c_1$ and finite
positive constants $c_2,\epsilon>0$, then also the full potential
including the fermionic fluctuations is bounded from below.

This result is in obvious direct disagreement with the standard
perturbative reasoning outlined above, cf. Eqs.~\eqref{eq:Ueff} and
\eqref{eq:betapert}. Nevertheless, it is in fact possible to
``rediscover'' this seeming instability of the standard reasoning from
the stable contribution \eqref{eq:SC1b} by trying to take the limit
$\Lambda\to \infty$. The leading-order terms in this limit read,
\begin{equation}
\begin{split}
U_{\text{F},a}(\rho) &=
-\frac{\Lambda^2}{8\pi^2} h_a^2 \rho
+ \frac{1}{16\pi^2} \left[ h_a^4 \rho^2 \ln \frac{\Lambda^2}{h_a^2\rho} \right. \\
&\hspace{75pt} \left. + \frac{h_a^4\rho^2}{2} + \mathcal{O}\left( \frac{(h_a^2\rho)^3}{\Lambda^2} \right) \right]. 
\end{split}
\label{eq:Ueffexp1}
\end{equation}
From here, it is tempting to isolate the divergencies $\sim \Lambda^2$
and $\ln \Lambda$, combine them with the bare scalar mass and $\phi^4$
coupling parameters, and trade them for renormalized parameters
$m_\phi^2(\mu_0)$ and $\lambda(\mu_0)$. Here $\mu_0$ is some arbitrary
(typically low-energy renormalization scale). Ignoring the mass term
for a moment, the renormalized interaction contribution to the
effective potential would then read
\begin{equation}
U_{\text{F},a}(\rho) \stackrel{\text{?}}{\to} 
- \frac{1}{16\pi^2} 
h_a^4 \rho^2 \left( \ln \frac{h_a^2\rho}{\mu_0^2}  +\text{const.} \right),
 \label{eq:Ueffexp2}
\end{equation}
where the constant depends on the details of the renormalization
scheme. This is precisely the fermion-loop contribution to the
  effective action, which we would obtain from integrating the first
  two terms of the $\beta_\lambda$ function \eqref{eq:betapert} from
  $\mu_0$ to $\mu$ and identifying $\mu^2 \sim \rho$. Hence, we have
  ``rederived'' the contribution with the characteristic minus sign
  that seems to indicate the presence of an instability at large
  values of $\rho$, while the cutoff $\Lambda$ seems to have
  disappeared completely.

The problem of this line of argument becomes obvious, once we go back
to the cutoff-dependent leading order terms in
\Eqref{eq:Ueffexp1}. It is straightforward to work out that also these
leading-order terms seem to have an instability: the interaction part
of the potential in square brackets first develops a maximum and then
eventually turns negative for large fields $\rho$. However, the
location of the maximum is in fact at $h_a^2\rho= \Lambda^2$. In other
words, these seeming instability features appear precisely at those
field values, where the expansion in terms of the parameter
$\frac{h_a^2\rho}{\Lambda^2}\ll 1$ breaks down. We conclude that the
instability ``discovered'' in \Eqref{eq:Ueffexp2} is an artifact of
having tried to send the cutoff to infinity $\Lambda\to \infty$
together with a problematic choice of renormalization conditions. In
fact, it has been shown in \cite{Branchina:2005tu,Gneiting:2005} for
the $\mathbb{Z}_2$-Yukawa model that the renormalization conditions
needed to arrive at \Eqref{eq:Ueffexp2} require an unstable bare
bosonic potential with negative bare $\phi^4$ coupling,
$\lambda(\Lambda)<0$.

Some additional comments are in order:

(1) Our conclusions are identical to those of
\cite{Holland:2003jr,Holland:2004sd}, where essentially the same
results have been found for the $\mathbb{Z}_2$-Yukawa model. In these
works, fully nonperturbative lattice simulations have been compared
with the one-loop effective potential with a cutoff kept finite,
matching the lattice data almost perfectly. By contrast, the effective
potential with the cutoff removed a la \Eqref{eq:Ueffexp2} shows an
artificial instability in strong disagreement with the nonperturbative
simulation. This work has been criticized
\cite{Espinosa:2013lma,Einhorn:2007rv} also because it is generically
difficult on the lattice to bridge wide ranges of scales, in
particular to separate the cutoff from the long-range mass scales by
many orders of magnitude. As is clear from the above discussion, this
problem does not exist for the present line of argument; the cutoff
can be arbitrarily large in the above discussion of the fermion
determinant. As long as it is finite, the interaction part of the
determinant does not induce any instability. 

(2) For the above discussion and the comparison to the standard line
of arguments at one-loop order, it has been sufficient to evaluate the
determinant for a homogeneous mean field. Though this does not interfere
with our argument, one might ask whether the determinant behaves
qualitatively differently for non-homogeneous fields. Some exact
results are known for $d=1+1$ dimensional determinants, where the
Peierls instability at a finite chemical potential can lead to
inhomogeneous ground states with lower free energy
\cite{Thies:2006ti}. However, the vacuum ground state is generically
homogeneous as no mechanism exists that can ``pay'' for the higher
cost in kinetic energy.  Absolute lower and upper bounds for fermion
determinants have been found, e.g., for QED \cite{Fry:2011iz}.

(3) The fact that the interaction part of the fermion contribution to
the scalar potential is positive does not imply that the full theory
cannot have further potentially (meta-)stable vacua. The conclusion
rather is that such further vacua have to be provided by the bosonic
sector. In particular, the bare bosonic potential $U_{\text{B}}$ can
in principle be chosen such that it has several vacua. As a special
case, it is even possible to construct somewhat special examples such
that the bare bosonic potential has one minimum, but the sum of
$U_{\text{B}}$ and $U_{\text{F}}$ has two minima. This is still very
different from the perturbative reasoning which for the present model
seems to suggest a global instability due to the fermionic
fluctuations, whereas a global instability of
$U_{\text{B}}+U_{\text{F}}$ in our analysis would have to be seeded
from the choice of $U_{\text{B}}$. In other words, our arguments do
not exclude that our electroweak vacuum is unstable, but they suggest
that such an in/meta-stability would have to be provided by the
microscopic underlying theory, see \cite{Hebecker:2013lha} for a
specific example from string phenomenology. In this case, however, the
Higgs mass bounds from metastability as well as the life-time
estimates of the electroweak vacuum would be very different from the
conventional estimates, see e.g. \cite{Branchina:2013jra}.

(4) As mentioned above, the result for the fermion determinant is
regulator dependent, as long as the cutoff is kept finite. The
preceding results have been derived for a sharp cutoff in momentum
space. These results in fact generalize to arbitrary smooth cutoff
shape functions in momentum space as they can be implemented
straightforwardly within the functional RG framework, see
below and App.~\ref{sec:MF}. Though this is not an issue for the present model, one might be
concerned about the fact that such regularizations are not
gauge invariant. Hence a gauge-invariant regularization is studied in
the remainder of this section.

\subsubsection{$\zeta$ function regularization}

It is illustrative to study the fermion determinant also using $\zeta$
function regularization which can be used to interpolate between
propertime and dimensional regularization. For this, we write
$U_{\text{F}}$ of \Eqref{eq:UF2} for one of the quark flavors as
\begin{equation}
U_{\text{F},a}(\rho) =  \frac{1}{2\Omega}
\int_{1/\Lambda^2}^\infty \frac{dT}{T} \left(e^{-h_a^2 \rho T}-1 \right)\, \text{Tr}\, e^{\partial^2 T},
\label{eq:UF1zeta}
\end{equation}
where $T$ is a propertime parameter, being introduced via Frullani's
formula for a representation of the logarithm. Here the lower bound of
the $T$ integral serves as a (gauge-invariant) momentum
cutoff. Furthermore, we now evaluate the momentum trace in $d$
dimensions and introduce an arbitrary dimensionful scale $\mu_0$ in
order to implement the correct dimensionality of the potential,
$\text{Tr} \to \text{tr}_\gamma\Omega \int \frac{d^4 p}{(2\pi)^4} \to
 \text{tr}_\gamma\frac{\Omega}{\mu_0^{d-4}} \int \frac{d^d p}{(2\pi)^d}$. We obtain,
\begin{equation}
U_{\text{F},a}(\rho) =   \frac{2\mu_0^{4-d}}{(4\pi)^{d/2}} 
\int_{1/\Lambda^2}^\infty \frac{dT}{T^{1+(d/2)}} \left(e^{-h_a^2 \rho T}-1 \right).
\label{eq:UF2zeta}
\end{equation}
In the limit $d\to4$, we have the standard propertime regularization,
whereas in the limit $\Lambda\to\infty$, we end up with dimensional
regularization. Separating the mass term $\sim\rho$ as before, we get
\begin{eqnarray}
U_{\text{F},a}(\rho) &=& - \frac{4\mu_0^{4-d}}{(d-2)(4\pi)^{d/2}} h_a^2 \rho \Lambda^{d-2}\label{eq:UF3zeta}\\
&&\!\!\! +\frac{2\mu_0^{4-d}}{(4\pi)^{d/2}} 
\!\!\int_{\!1/\Lambda^2}^\infty\!\! \frac{dT}{T^{1+(d/2)}} \left(e^{-h_a^2 \rho T} +h_a^2\rho T -1 \right).\nonumber\\
&&
\label{eq:UF4zeta} 
\end{eqnarray}
Again, we observe that the mass term (first line) contributes with the
minus sign as expected, whereas the interaction part (second line) is
a strictly positive function for all finite values of
$\Lambda,h_a,\rho,d>0$. The conclusions are therefore identical to the
ones for the sharp cutoff. Incidentally, the propertime integration
can be carried out analytically, the result can be written in terms of
an incomplete $\Gamma$ function with the positivity properties of
course remaining unchanged.

As the mass term will become part of the renormalized scalar mass term
by means of a renormalization condition, let us now focus on the
interaction part \Eqref{eq:UF4zeta}. In order to take the limit
towards dimensional regularization, we first take the limit
$\Lambda\to\infty$, and then expand about $d=4-\epsilon$, as is
standard. We then find for the interaction part
\begin{eqnarray}
U_{\text{F},a}(\rho) &\to& \frac{2\mu_0^{4-d}}{(4\pi)^{d/2}} (h_a^2 \rho)^{d/2} \Gamma(-d/2)\label{eq:UF5zeta}\\
&=& \frac{h_a^4 \rho^2}{8\pi^2} \frac{1}{\epsilon} \nonumber\\
&&- \frac{h_a^4 \rho^2}{16\pi^2}\left( \ln \frac{h_a^2 \rho}{\mu_0^2} + \text{const.}\right) + \mathcal{O}(\epsilon).
\label{eq:UF6zeta} 
\end{eqnarray}
Following the standard recipes, we would absorb the positive $1/\epsilon$
divergence in the bare $\phi^4$ term by means of a renormalization
condition. The finite part in \Eqref{eq:UF6zeta} is identical to that
of the sharp cutoff in \Eqref{eq:Ueffexp2} in the limit
$\Lambda\to\infty$ (apart from the scheme-dependent constants),
seemingly indicating an instability at large field values. With
dimensional regularization, we would therefore arrive at the standard
conclusion that fermionic fluctuations can induce an instability of
the vacuum at large fields.

Whereas for the sharp cutoff, this was an obvious artifact of the
$\Lambda\to\infty$ limit, the failure is less obvious
here. Nevertheless, as we have derived this misleading result of a
negative contribution from a strictly positive expression given in
\Eqref{eq:UF4zeta}, it is clear that the standard strategies of
dimensional regularization fail to describe the global behavior of the
fermion determinant properly. The reason is that dimensional
regularization is not only a regularization but at the same time a
projection solely onto the logarithmic divergencies.  It has in fact
long been known that the use of dimensional regularization in the
presence of large fields can become delicate; procedures to deal with
this problem typically suggest to go back to the dimensionally
continued propertime/$\zeta$-function representation that we started
out with \cite{Brown:1976wc,Luscher:1982wf}.

There is another perspective that explains why the standard
perturbative argument of integrating the $\beta$ function of the
$\phi^4$ theory is misleading as far as vacuum stability is concerned:
the $\beta$ functions are typically derived in mass-independent
regularization schemes (though mass-dependent schemes have recently
also been studied \cite{Spencer-Smith:2014woa}), and it is implicitly
assumed that the discussion can be performed in the deep Euclidean
region where all mass scales are much smaller than any of the involved
momentum scales of the fluctuations. The latter assumption is in fact
not valid, as both scales the value of the field as well as the cutoff
$\Lambda$ can interfere non-trivially with each other. This is
illustrated rather explicitly in the sharp-cutoff calculation given
above.

\section{Renormalization flow}
\label{sec:flow}

Independently of the validity of the perturbative arguments about
vacuum stability, the comparatively small mass of the observed Higgs
boson \cite{Aad:2012tfa} poses a challenge: the fermionic fluctuations
(dominated by top loops) contribute to the curvature of the effective
potential which determines the Higgs boson mass. Even in the absence
of any bosonic interactions, this appears to lead to a lower bound on
the value of the Higgs mass which are in tension with the measured
value. This line of argument has been used in quantitative lattice
studies \cite{Fodor:2007fn,Gerhold:2007yb}. For rendering simulations
on a Euclidean lattice well defined, the bosonic action has
to be bounded from below; in practice, lower bounds on the Higgs mass
thus arise in the limit of the bare $\phi^4$ coupling approaching
zero, $\bar\lambda\to 0_+$.

While it may be debatable whether this criterion could be relaxed for
a Minkowskian functional integral in the continuum, we have already
provided first examples in the $\mathbb{Z}_2$-symmetric Yukawa toy
model, that these conventional lower-bounds can be relaxed by allowing
for more general forms of the bare potential
\cite{Gies:2013fua}. These results have recently been confirmed in
lattice simulations \cite{Hegde:2013mks}. In particular, no state of
meta-/instability is required for relaxing the lower bound. In the
following, we generalize these results to the chiral top-bottom-Higgs
Yukawa model.

For this, we use the functional RG as a nonperturbative tool. Though
our results for the lower Higgs mass bound can in principle also be
derived within a perturbative framework, the functional RG allows to
discuss weak-coupling limits (lower bounds) as well as the
large-coupling region (upper bounds) in a unified setting. Also, more
generalized bare actions can be dealt with more conveniently, the
preceding fermion determinant results arise in a specific limit, and
higher-loop effects as well as RG improvement are automatically
included. Starting from a bare microscopic action $S$ defined at a UV
cutoff $\Lambda$ (which may or may not be sent to infinity), the RG flow of a corresponding scale-dependent action $\Gamma_k$ is determined by the Wetterich equation
\cite{Wetterich:1992yh}
\begin{align}
 \pt \Gamma_k = \frac{1}{2} \STr \Big[ \frac{\pt R_k}{\Gamma_k^{(2)}+R_k} \Big], \quad t=\ln\frac{k}{\Lambda},
 \label{WetterichEq}
\end{align}
which interpolates smoothly between microscopic physics,
$\Gamma_{k=\Lambda}=S$ and the full quantum effective action
$\Gamma_{k=0}=\Gamma$. $\Gamma_k^{(2)}$ in \Eqref{WetterichEq} denotes
the second derivative of $\Gamma_k$ with respect to the fluctuating
fields ($\phi_a$, $t_{\text{L}}$, $\bar{t}_{\text{L}}$, $\cdots$) and the supertrace
includes a minus sign for fermionic degrees of freedom. The regulator
$R_k$ in the denominator acts as an IR cutoff for modes with momenta
smaller than $k$. The derivative $\pt R_k$ at the same time provides
for a UV regularization. For instance, for a suitable choice of $R_k$
in terms of a sharp cutoff, taking only the fermion loops of
\Eqref{WetterichEq} into account reproduces the fermion determinant
results given above. For detailed reviews of the functional RG,
see~\cite{Berges:2000ew,Aoki:2000wm,Pawlowski:2005xe,Gies:2006wv,
  Delamotte:2007pf,Kopietz:2010zz,Braun:2011pp}.

We compute the effective average action at next-to-leading order (NLO)
in a derivative expansion, corresponding to the following truncation:
\begin{align}
 \Gamma_k = \int_x 
 &\Big[ Z_{\phi}|\partial_{\mu}\phi|^2 + U(\phi^{\dagger}\phi) + Z_{\text{L}}\bar{\psi}_{\text{L}}i\slashed{\partial}\psi_{\text{L}} + Z_{\text{R}}^{\text{t}}\bar{t}_{\text{R}}i\slashed{\partial}t_{\text{R}} \notag \\
 &+ Z_{\text{R}}^{\text{b}}\bar{b}_{\text{R}}i\slashed{\partial}b_{\text{R}} + i\bar{h}_{\text{b}} (\bar{\psi}_{\text{L}}\phi b_{\text{R}} + \bar{b}_{\text{R}}\phi^{\dagger}\psi_{\text{L}}) \label{truncation} \\
 &+ i\bar{h}_{\text{t}} (\bar{\psi}_{\text{L}}\phi_{\mathcal{C}} t_{\text{R}} + \bar{t}_{\text{R}}\phi_{\mathcal{C}}^{\dagger}\psi_{\text{L}}) \Big]. \notag
\end{align}
Here the scalar potential, both Yukawa couplings and the wave function
renormalizations $Z_{\phi}$, $Z_{\text{L}}$ and $Z_{\text{R}}^{\text{t,b}}$ for 
the fields depend on the RG scale $k$ ($U=U_k$, $\htop=h_{\text{t},k}$,
$\cdots$). For compactness of notation, this dependence is suppressed.

Inserting this truncation into the Wetterich equation leads to the $\beta$
functions, i.e.,  the flow equations for the effective potential,
the Yukawa couplings as well as for the wave function
renormalizations. The latter will be encoded in the anomalous
dimensions of the fields,
\begin{align*}
 \eta_i = -\pt \ln{Z_i},
\end{align*}
where $i$ labels the different fields. Furthermore, it is useful to
define dimensionless renormalized quantities, such as
\begin{align*}
 \rho &= Z_{\phi}k^{2-d}\phi^{\dagger}\phi \\
 \htop^2 &= Z_{\phi}^{-1}Z_{\text{L}}^{-1}{Z_{\text{R}}^{\text{t}}}^{-1}k^{d-4}\, \bar{h}_{\text{t}}^2, \\
 \hbot^2 &= Z_{\phi}^{-1}Z_{\text{L}}^{-1}{Z_{\text{R}}^{\text{b}}}^{-1}k^{d-4}\, \bar{h}_{\text{b}}^2.
\end{align*}
Here and in the following, we work in $d$ dimensional spacetime for
reasons of generality.  Accordingly, the dimensionless potential
simply reads:
\begin{align*}
 u = k^{-d} U.
\end{align*}
In the following, we list the flow equations for the various
quantities, as they follow from the Wetterich equation, using standard
calculation techniques, see, e.g., \cite{Jungnickel:1995fp}. The flow
equation for the potential can be written as:
\begin{equation}
\begin{split}
 \pt u &= -du + (d-2+\eta_{\phi})\rho u' \\
 &\quad + v_d \Big( 3l_0^d(u') + l_0^d(u'+2\rho u'') \Big) \\ 
 &\quad - d_{\text{W}}v_d\Big( l_{0\, \text{L}}^{(\text{F})d}(\htop^2\rho) + l_{0\, \text{L}}^{(\text{F})d}(\hbot^2\rho) \\
 &\qquad\qquad\quad + l_{0\, \text{R}_{\text{t}}}^{(\text{F})d}(\htop^2\rho) + l_{0\, \text{R}_{\text{b}}}^{(\text{F})d}(\hbot^2\rho) \Big),
\end{split}
 \label{flowpot}
\end{equation}
where primes denote derivatives with respect to $\rho$. The threshold
functions $l_0^d$, $l_{0\, \text{L}}^{(\text{F})d}$, $l_{0\,
  \text{R}_{\text{t}}}^{(\text{F})d}$ and $l_{0\,
  \text{R}_{\text{b}}}^{(\text{F})d}$ governing the decoupling of
massive modes can be found in the appendix for the convenient choice
of a linear regulator. Here, $v_d^{-1}=2^{d+1}\pi^{d/2}\Gamma(d/2)$
and $d_{\text{W}}$ is the dimension of the spinor representation of
the Weyl fermions. We will specialize to $d=4$ and $d_{\text{W}}=2$
for quantitative calculations. The flow equations for the Yukawa
couplings are
\begin{widetext}
\begin{equation}
\begin{split}
 \pt \htop^2 &= (d-4+\eta_{\phi}+\eta_{\text{L}}+\eta_{\text{R}}^{\text{t}})\htop^2 \\
 &\quad - 4v_d \htop^4 \Big[
 (6\kappa u''+4\kappa^2u''') l_{1,2}^{(\text{FB})d}(\htop^2\kappa,u'+2\kappa u'') - 2\kappa u'' l_{1,2}^{(\text{FB})d}(\htop^2\kappa,u') \\
 &\quad + 2\htop^2\kappa\big( l_{2,1}^{(\text{FB})d}(\htop^2\kappa,u'+2\kappa u'') - l_{2,1}^{(\text{FB})d}(\htop^2\kappa,u') \big) - l_{1,1}^{(\text{FB})d}(\htop^2\kappa,u'+2\kappa u'') + l_{1,1}^{(\text{FB})d}(\htop^2\kappa,u') \Big] \\
 &\quad - 8v_d \htop^2\hbot^2 \Big[ - 2\kappa u'' l_{1,2}^{(\text{FB})d}(\hbot^2\kappa,u')
 - 2\hbot^2\kappa\, l_{2,1}^{(\text{FB})d}(\hbot^2\kappa,u')
 + \, l_{1,1}^{(\text{FB})d}(\hbot^2\kappa,u') \Big] \Big|_{\rho=\kappa},
 \\
 \pt \hbot^2 &= (d-4+\eta_{\phi}+\eta_{\text{L}}+\eta_{\text{R}}^{\text{b}})\hbot^2 \\
 &\quad - 4v_d \hbot^4 \Big[
 (6\kappa u''+4\kappa^2u''') l_{1,2}^{(\text{FB})d}(\hbot^2\kappa,u'+2\kappa u'') - 2\kappa u'' l_{1,2}^{(\text{FB})d}(\hbot^2\kappa,u') \\
 &\quad + 2\hbot^2\kappa\big( l_{2,1}^{(\text{FB})d}(\hbot^2\kappa,u'+2\kappa u'') - l_{2,1}^{(\text{FB})d}(\hbot^2\kappa,u') \big) - l_{1,1}^{(\text{FB})d}(\hbot^2\kappa,u'+2\kappa u'') + l_{1,1}^{(\text{FB})d}(\hbot^2\kappa,u') \Big] \\
 &\quad - 8v_d \hbot^2\htop^2 \Big[ - 2\kappa u'' l_{1,2}^{(\text{FB})d}(\htop^2\kappa,u')
 - 2\htop^2\kappa\, l_{2,1}^{(\text{FB})d}(\htop^2\kappa,u')
 + \, l_{1,1}^{(\text{FB})d}(\htop^2\kappa,u') \Big] \Big|_{\rho=\kappa},
 \end{split}
 \label{flowYukawa}
\end{equation}
where $\kappa$ denotes the minimum of the potential. Finally, the anomalous dimensions are given by
\begin{equation}
\begin{split}
\eta_{\phi} &= \frac{8v_d}{d}\kappa \big[ 3u''m_{22}^d(u') + (3u''+2\kappa u''')^2 m_{22}^d(u'+2\kappa u'') \big] \\
 &\quad -\frac{8d_{\text{W}}}{d}v_d \Big[ \kappa \htop^4 m_2^{(\text{F})d}(\htop^2\kappa) - \htop^2 m_4^{(\text{F})d}(\htop^2\kappa;\eta_{\text{t}}) + \kappa \hbot^4 m_2^{(\text{F})d}(\hbot^2\kappa) - \hbot^2 m_4^{(\text{F})d}(\hbot^2\kappa;\eta_{\text{b}}) \Big] \Big|_{\rho=\kappa},
 \\
 \eta_{L} &= \frac{4v_d}{d} \Big[ \htop^2\Big( m_{1,2}^{(\text{FB})d}(\kappa \htop^2,u') + m_{1,2}^{(\text{FB})d}(\kappa \htop^2,u'+2\kappa u'') \Big) + 2\hbot^2 m_{1,2}^{(\text{FB})d}(\kappa \hbot^2,u') \Big] \Big|_{\rho=\kappa},
 \\
 \eta_{\text{R}}^{\text{t}} &= \frac{4v_d}{d}\htop^2 \big[ m_{1,2}^{(\text{FB})d}(\htop^2\kappa,u')+m_{1,2}^{(\text{FB})d}(\htop^2\kappa,u'+2\kappa u'') + 2m_{1,2}^{(\text{FB})d}(\hbot^2\kappa,u') \big] \Big|_{\rho=\kappa},
 \\
 \eta_{\text{R}}^{\text{b}} &= \frac{4v_d}{d}\hbot^2 \big[ m_{1,2}^{(\text{FB})d}(\hbot^2\kappa,u')+m_{1,2}^{(\text{FB})d}(\hbot^2\kappa,u'+2\kappa u'') + 2m_{1,2}^{(\text{FB})d}(\htop^2\kappa,u') \big] \Big|_{\rho=\kappa}.
\end{split}
\label{eq:anomalous}
\end{equation}
\end{widetext}
Again, the definition of the threshold functions can be found in the
appendix. Furthermore it is useful to define anomalous dimensions of
the top and bottom quark via
\begin{align}
 \eta_{\text{t}} = \frac{1}{2} (\eta_{\text{L}} + \eta_{\text{R}}^{\text{t}}), \quad \eta_{\text{b}} = \frac{1}{2} (\eta_{\text{L}} + \eta_{\text{R}}^{\text{b}}).
 \label{eq:anomalous2}
\end{align}
The flow equations agree with those for the $\mathbb{Z}_2$-symmetric
Yukawa model \cite{Gies:2013fua,Gies:2009hq} in the limit of a
vanishing bottom sector, $\hbot=0$, and ignoring the terms arising
from the additional scalar contributions.\footnote{As further
  cross-checks, we note that our flows also agree with those of
  \cite{Gies:2009sv,Gies:2009da} for $N_{\text{L}}=2$ and $\hbot=0$
  (apart from a missing factor of $1/2$ in $\eta_{\text{L,R}}$ as also
  noted in \cite{Janssen:2012oca}). We also observe agreement with the
  flows of \cite{Gies:2013pma} for $\hbot=0$ and upon dropping the
  gauge sector in that work.}  The reliability of the derivative 
expansion can be monitored with the aid of the anomalous dimension,
providing a rough measure for the importance of the higher-derivative
terms. In practice, we study the convergence of the derivative
expansion by comparing leading-order results ($\eta_i=0$) to the full
NLO calculation, see below.

%
%
%
%

\section{Nonperturbative Higgs mass bounds}
\label{sec:mHbounds}

The flow equations listed above enable us to take a fresh look at
Higgs boson mass bounds possibly arising from the RG flow of the
model. For this, it is useful to think of the RG flow as a mapping
from a microscopic theory defined at some high scale $\Lambda$ onto
the effective long-range theory governing the physics observed in
collider experiments. For this mapping, we use the standard-model-type
parameters $v=246$GeV, $\mtop=173$GeV and $\mbot=4.2$GeV as
constraints. The range of all possible Higgs boson masses resulting
from the remaining UV parameters for a given cutoff then defines the
IR window and correspondingly puts bounds on the Higgs boson mass as a
function of the cutoff scale $\Lambda$. 

In full generality, constructing this mapping is a complex problem,
since the microscopic theory at scale $\Lambda$ is a priori
unconstrained to a large extent. At first sight, it seems natural to
allow only the renormalizable terms in the bare action. For the
present model, this has been successfully implemented in extensive
lattice simulations \cite{Fodor:2007fn,Gerhold:2007yb, Gerhold:2010wv}
yielding quantitative results for the Higgs mass bounds. In
particular, the lower bound arises from the lowest possible value for
the Higgs selfinteraction $\bar{\lambda}\phi^4$, i.e.,
$\bar\lambda\to0$, for which the lattice theory remains
well-defined. The resulting bounds should therefore not be
viewed as vacuum stability bounds, but as consistency bounds arising
from the requirement that the underlying lattice partition function
is well defined.

However, there is no need at all to confine the bare theory to just
the renormalizable operators. On the contrary, generic underlying
theories (UV completions) are expected to produce all terms allowed by
the symmetries, such that the search for Higgs mass bounds corresponds
to finding extrema of a function (the Higgs mass) depending on
infinitely many variables (the bare action). Formulated in terms of
this generality, it is actually unclear whether these extrema
and thus a universal consistency bound exist at all. We
therefore confine our study to a much simpler question: given the
lower mass bound arising within the conventional class of $\phi^4$
potentials, can we find more general bare potentials, which (a) lower
the mass bounds and (b) do not show an instability towards a different
vacuum neither in the UV nor in the IR? While (a) is obviously
inspired by the fact that the measured Higgs boson mass seems to lie
below the conventional lower bound, an answer to (b) can serve as an
illustration that no meta-stability is required in order to relax the
lower bound.

While these two questions have been answered in the affirmative for
the $\mathbb{Z}_2$-Yukawa model in \cite{Gies:2013fua} with functional
RG methods as well as for the present chiral model with lattice
simulations up to cutoff scales on the order of several TeV
\cite{Hegde:2013mks}, the present flow equation study can elucidate
the underlying RG mechanisms in more detail and can bridge a wide
range of scales. Furthermore, it is straightforward to deal with two
distinct quark masses, $\htop \neq \hbot$, in our functional approach,
whereas simulations with the physical ratio $\mtop/\mbot \simeq 40$
would be rather expensive on the lattice.

Before we turn to a quantitative analysis of the RG flow, we have to
cure a deficiency of the chiral Yukawa model in comparison with the
full standard model top-bottom-Higgs sector. Since chiral symmetry breaking
in the present model breaks a global symmetry, our present model has
massless Goldstone bosons in the physical spectrum. This is different
from the standard model where the would-be Goldstone bosons due to
their interplay with the gauge sector are ultimately absent from the
physical spectrum, the latter finally containing massive vector
excitations. In order to make contact with the standard model physics,
we therefore have to modify our chiral Yukawa model, otherwise the
massless Goldstone modes have the potential to induce an IR behavior
which is very different from that of the standard model. This
modification of the model is not unique and could be done in various
ways. For instance on the lattice, the influence of the unwanted
Goldstone bosons is identified by their strong finite volume effects
and subtracted accordingly \cite{Gerhold:2007yb,
  Gerhold:2010wv}. Similarly, we could study the onset of Goldstone
dynamics in the limit $k\to 0$ and perform a similar subtraction.

In the present work, we model the decoupling of Goldstone bosons more
physically as inspired by the Higgs mechanism in the fully gauged
version of the theory: generically all dependencies on the particle
masses and their decoupling is contained in the threshold functions of
the flow equations. For the linear regulators used below, this
dependence occurs in the form
\begin{align}
 \frac{k^2}{k^2+m^2}
 \label{genericcontributiontoflow}
\end{align}
to some power. For $k\to0$, all these functions vanish for finite
particle masses $m$, whereas massless modes such as Goldstone modes
with $m=m_{\text{G}}=0$ contribute equally on all scales $k$.  (For
other regulators, the functional dependence on $k$ and $m$ may look
differently, but behaves analogously in the various limits.) The
Goldstone modes can therefore directly be identified in our flow
equations. They contribute to those threshold functions that contain
an argument $\sim u'(\rho)$ in the SSB regime. As soon as we enter the broken regime,
the corresponding mass argument $m_{\text{G}}^2/k^2 \sim u'(\kappa)$
vanishes at the running minimum $\kappa$ of the potential. We thus can
dynamically remove the Goldstone modes by the replacement
\begin{align*}
 \frac{k^2}{k^2+m_{\text{G}}^2} \rightarrow \frac{k^2}{k^2+m_{\text{G}}^2 +
   g\frac{v_k^2}{2}} \stackrel{m_{\text{G}}=0}{=} \frac{k^2}{k^2 +
   g\frac{v_k^2}{2}},
\end{align*}
in the broken regime. Here, $v_k$ is the running vacuum expectation
value approaching $v_k\to v$ in the long range limit $k\to0$, and $g$
is an a priori free parameter. Inspired by the decoupling of massive
vector bosons in the full standard model, we choose $g$ in such a way
that the resulting masses for the Goldstone bosons have the same order
of magnitude as the $W$ boson mass scale, e.g. $g=2(80/246)^2$. It
turns out that the results for the lower Higgs mass bound is only slightly
affected by different choices of $g$, as well as different choices of
removing the Goldstone mode contributions,
cf. App.~\ref{CutOffGoldstones}.

As a result, all fluctuations acquire a mass in the regime of
spontaneous symmetry breaking (SSB) and the whole flow freezes out,
similarly to the $\mathbb{Z}_2$ Yukawa model and as expected in the
full standard model.

Now that we have amended our model with a dynamical removal of the
unwanted Goldstone bosons, the RG flow of the model is technically
similar to the simpler $\mathbb{Z}_2$ invariant Yukawa model
extensively studied in \cite{Gies:2013fua}. In the following, we
therefore focus on the new features induced by the additional degrees
of freedom of the chiral model such as the bottom quark and the three
additional real scalar fields. Further technical details follow those
of \cite{Gies:2013fua}.

The relevant information to compute Higgs mass bounds can be extracted
from the shape of the scalar effective potential near its minimum.  We
extract this information from a power series expansion of the
potential about this flowing minimum.  In the symmetric (SYM)
regime, we expand about zero field-amplitude
\begin{equation*}
 u = \sum_{n=1}^{\Np} \frac{\lambda_n}{n!} \rho^n
\end{equation*}
in which $\lambda_1$ is a mass term and $\lambda_2$ the quartic
coupling. $\Np$ defines the order of this polynomial expansion. In
practice, all results studied in this work converge rather rapidly in
this expansion.  The expansion point in the SSB regime is set by the
nonvanishing vev
\begin{equation*}
 u = \sum_{n=2}^{\Np} \frac{\lambda_n}{n!} (\rho-\kappa)^n.
\end{equation*}
The flow equations for the couplings $\lambda_1,\cdots,\lambda_{N_P}$
(SYM) or $\kappa$, $\lambda_2,\cdots,\lambda_{N_P}$ (SSB) can be read
off from Eq. (\ref{flowpot}).

\subsection{Bare potentials of $\phi^4$ type}

First, we determine mass bounds for the Higgs boson arising from
microscopic bare potentials of $\phi^4$ type,
\begin{align}
 u_{\Lambda} &= \lambda_{1,\Lambda}\rho + \frac{\lambda_{2,\Lambda}}{2}\rho^2 &\mathrm{(SYM)}
 \\
 u_{\Lambda} &= \frac{\lambda_{2,\Lambda}}{2}(\rho-\kappa_{\Lambda})^2  &\mathrm{(SSB)}.
\end{align}
For small $\lambda_{2,\Lambda}$ a physical flow typically starts in
the SYM regime. Near the electroweak scale the system is driven into
the SSB regime by fermionic fluctuations. Mathematically speaking, we
switch from the flow eq. for the SYM to the SSB couplings at the
scale, where $\lambda_1$ crosses zero. In the SSB regime a nonzero vev
builds up, inducing masses for all particles in the theory including
the would-be Goldstone bosons as discussed above. This results in a
decoupling of all modes in the IR and therefore all dimensionful
quantities freeze out.

By contrast: for large $\lambda_{2,\Lambda}$, the theory already
starts in the SSB regime with a small value for
$\kappa_{\Lambda}$. The flow still runs over many scales, depending on
the initial conditions, until $\kappa$ eventually grows large near the
electroweak scale. As a result, all modes decouple and we can read off
the long-range observables.

The flow equation provide us with a map of the UV parameters to physical
parameters such as the mass of the Higgs, the top or the bottom
quark. In the following, we fine tune either $\lambda_{1,\Lambda}$ if
we start in the SYM regime or $\kappa_{\Lambda}$ in the SSB regime, in
order to arrive at a vev of $v_{k\to 0}=246$GeV in the IR. Further,
we vary the bare top $h_{t,\Lambda}$ and bottom $h_{b,\Lambda}$ Yukawa
coupling such that we obtain the desired top and bottom quark mass,
$\mtop \simeq 173$GeV and $\mbot \simeq 4.2$GeV.  For this reduced
class of bare $\phi^4$ potentials, the Higgs mass is only a function of
the bare quartic coupling $\lambda_{2,\Lambda}$ for a fixed cutoff. In
order to start with a well defined theory in the UV,
$\lambda_{2,\Lambda}$ must be strictly nonnegative.

\begin{figure}
 \centering
 \includegraphics[width=8.5cm]{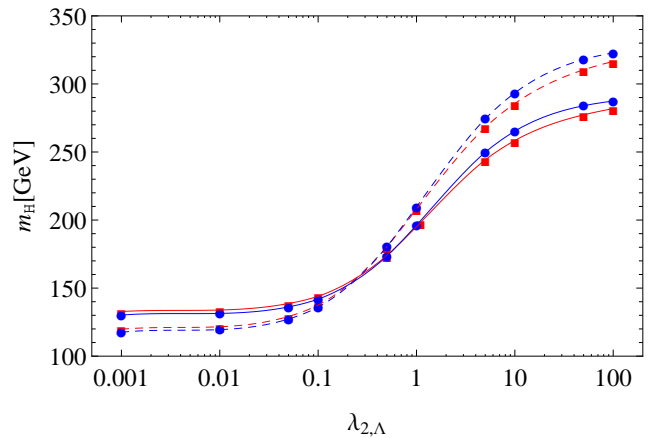}
 \caption{Higgs mass $\mH$ as a function of the bare quartic coupling
   $\lambda_{2,\Lambda}$ for fixed cutoff $\Lambda=10^7$GeV for
   various approximations. Dashed lines depict leading order results
   in the derivative expansion while solid lines show the next-to-leading
   order. Also the convergence of the polynomial truncation of the
   scalar potential is illustrated. Red lines with squares arise from
   $\Np=2$ whereas blue lines with circles are derived for
   $\Np=4$.}
 \label{fig:IRWindowHiggsMassWithinPhi4}
\end{figure}

We find that the Higgs mass is a monotonically increasing function of
the bare quartic coupling, which can be seen in
Fig.~\ref{fig:IRWindowHiggsMassWithinPhi4}. Here, the Higgs mass $\mH$
is plotted as a function of the bare quartic coupling
$\lambda_{2,\Lambda}$ for a fixed cutoff $\Lambda=10^7$GeV. The lower
bound is approached for $\lambda_{2,\Lambda}\to 0$, where the Higgs
mass becomes rather independent of $\lambda_{2,\Lambda}$. This was
also shown in lattice simulations for the $\mathbb{Z}_2$ model
\cite{Holland:2003jr} as well as for a chiral Yukawa theory
\cite{Gerhold:2007yb}. For large bare quartic couplings the Higgs mass
reaches a region of saturation.
To test the convergence of our expansion and truncation, we plotted
the Higgs mass in various approximations in
Fig.~\ref{fig:IRWindowHiggsMassWithinPhi4}. The derivative expansion
is tested by comparing leading-order (LO) (dashed lines) to NLO
results (solid lines).  At LO, we drop the running of the kinetic
terms in Eq. ($\ref{truncation}$), achieved by setting the anomalous
dimensions to zero in the flow equation ($\ref{flowpot}$) and
($\ref{flowYukawa}$).  These differ by at most $12\%$ for small as
well as for large couplings. The difference for small couplings is
somewhat larger than in the $\mathbb{Z}_2$-symmetric Yukawa model
because of the larger number of fluctuating scalar components.

Furthermore, we varied $\Np$ to check the convergence of the
polynomial expansion of the potential. The simplest nontrivial order
is given by $\Np=2$ and plotted as red lines with squares in
Fig.~\ref{fig:IRWindowHiggsMassWithinPhi4}. For $\Np=4$ (blue lines
with circles) there are only small deviations for small
$\lambda_{2,\Lambda}$ ($\sim 2$GeV) and deviations of $5\%$ for large
$\lambda_{2,\Lambda}$ compared to $\Np=2$. Beyond this, we find no
deviations between the Higgs masses for $\Np=4,5,6,8,10$ within our
numerical accuracy, demonstrating the remarkable convergence of the
polynomial truncation for the present purpose.

\begin{figure}
 \centering
 \includegraphics[width=8.5cm]{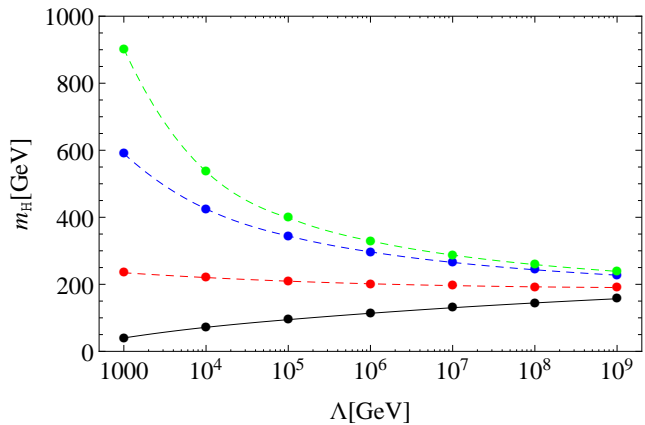}
 \caption{Higgs mass $\mH$ as a function of the cutoff $\Lambda$ for
   various bare quartic couplings. The black solid line represents a
   lower mass bound ($\lambda_{2,\Lambda}=0$) within $\phi^4$
   theory. Dashed lines depict Higgs masses for
   $\lambda_{2,\Lambda}=1$ (red), 10 (blue), 100 (green) from bottom
   to top.}
 \label{fig:HiggsMassBoundsWithinPhi4}
\end{figure}

In Fig.~\ref{fig:HiggsMassBoundsWithinPhi4}, resulting Higgs masses
are plotted as a function of the UV cutoff for different bare quartic
couplings for a wide range of cutoff values
$\Lambda=10^3,\cdots,10^9$GeV.  The lower black line is derived for
$\lambda_{2,\Lambda}=0$ and indicates a lower bound for $\mH$ within
the $\phi^4$ type bare potentials. Incidentally, it agrees
comparatively well with the results of a simple mean-field
(``large-$\Nf$'') calculation sketched in the Appendix. Dashed lines depict upper Higgs mass bounds if
one restricts the bare quartic coupling to $\lambda_{2,\Lambda}\leq
1,10,100$ (from bottom to top). Artificially restricting the coupling
$\lambda_{2,\Lambda}$ to the perturbatively accessible domain,
say, $\lambda_{2,\Lambda}\lesssim 1$, the upper bound is
obviously significantly underestimated.

\begin{figure}
 \centering
 \includegraphics[width=8.5cm]{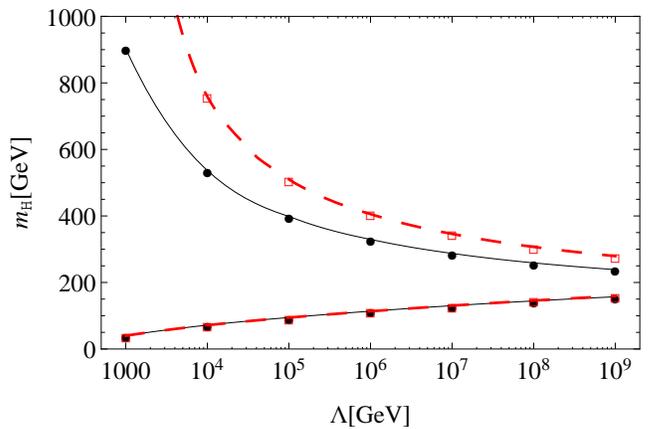}
 \caption{Higgs mass $\mH$ as a function of the cutoff $\Lambda$ for
   the chiral Higgs-Yukawa model (black solid lines) as well as for
   the simple $\mathbb{Z}_2$-symmetric Higgs-Yukawa theory (red dashed
   lines) as studied in \cite{Gies:2013fua}.  For the lower mass bound
   no significant difference is observed between the two models.  By
   contrast, a strongly coupled scalar sector
   ($\lambda_{\Lambda}=100$) leads to significantly lower masses in
   the present model which is a consequence of the additional scalar
   degrees of freedom in the chiral model, see main text.}
 \label{fig:Chiral-VS-Simple}
\end{figure}

By comparing the chiral Yukawa model to the $\mathbb{Z}_2$ Yukawa
model, we are able to study the influence of the additional standard
model degrees of freedom on the Higgs mass bounds.  As expected, the
bottom quark has no significant influence on the Higgs mass values,
due to its substantially smaller Yukawa coupling.  Higgs mass values
only differ by less than $1$ GeV if one neglects the coupling of the
bottom to the Higgs.  The main new contributions to the scalar
potential and thus to the Higgs mass are induced by the additional
scalar degrees of freedom.  For the lower bound
$\lambda_{2,\Lambda}=0$ the scalar sector is weakly coupled, hence the
Higgs mass is mainly build up by top fluctuations (apart from mutual
RG backreactions).  Therefore, the deviations between the two models
are small for the lower Higgs mass bounds.  For a strongly coupled
scalar sector in the UV, $\lambda_{2,\Lambda}>1$, the situation is
different.  There, the additional scalar degrees of freedom have a
significantly larger impact. This results in smaller Higgs masses,
since scalar fluctuations generically tend to drive the system into
the SYM regime.  The consequence of this is a flattening of the
scalar potential near its minimum and hence a smaller value for the Higgs mass  
which is visualized in Fig.~\ref{fig:Chiral-VS-Simple}.
%
%

Finally, we should emphasize once more that the use of
standard-model-like parameters is only for the purpose of
illustration. The quantitative difference becomes obvious, e.g., from
Fig.~\ref{fig:HiggsMassBoundsWithinPhi4} where the ``channel'' of
Higgs mass values that allow for a large cutoff is centered near
$\mH\simeq200$GeV. The same channel-like behavior in the full standard
model occurs near $\mH\simeq130$GeV. This quantitative difference is
mainly due to the influence of the gauge sectors, in particular the
strong interactions. But also the electroweak gauge sector can take a
conceptually (if not quantitatively) important influence on mass
bounds: e.g., recent nonperturbative lattice simulations of the
Yang-Mills-Higgs system suggest that the Higgs mass has to be larger
than the weak gauge boson masses in certain parameter regimes,
otherwise the electroweak sector would rather be in a QCD-like domain
\cite{Maas:2013aia}.

\subsection{Generalized bare potentials}

Motivated by previous continuum calculations in the $\mathbb{Z}_2$
model \cite{Gies:2013fua} and by lattice studies in the chiral version
\cite{Hegde:2013mks}, we study whether more general bare potentials
can modify the phenomenologically relevant lower Higgs mass bound. The
main purpose of this study is to demonstrate that the lower bound can
be relaxed without the occurrence of an in- or metastability of the
potential. An analysis of all conceivable bare potentials is a
numerically challenging problem and beyond the scope of this work.

In fact, already the simplest extension including a $\phi^6$ term in
the bare potential,
\begin{equation}
 u_{\Lambda} = \lambda_{1,\Lambda}\rho +
 \frac{\lambda_{2,\Lambda}}{2}\rho^2 +
 \frac{\lambda_{3,\Lambda}}{6}\rho^3,
\label{eq:genpot}
\end{equation}
suffices to illustrate our main point.  Here, negative values for the
bare quartic coupling $\lambda_{2,\Lambda}$ are permissible if the
potential is stabilized by a positive $\lambda_{3,\Lambda}$.
Precisely choices of this type indeed lead to the desired features.

\begin{figure}
 \centering
 \includegraphics[width=8.5cm]{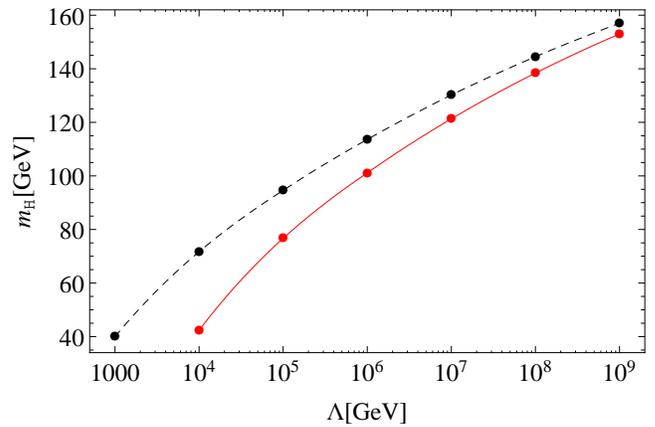}
 \caption{Higgs mass $\mH$ as a function of the cutoff $\Lambda$. The
   black dashed curve again corresponds to the lower bound derived
   within the class of $\phi^4$-type bare potential. The red solid
   line shows an example of Higgs boson mass values derived from a
   more general class of bare potentials of \Eqref{eq:genpot} with the
   initial UV values $\lambda_{2,\Lambda}=-0.1$ and
   $\lambda_{3,\Lambda}=3$. This demonstrates both that the lower
   bound can be significantly relaxed as well as that no in- or
   metastability is required to occur for Higgs masses below the
   conventional lower bound.}
 \label{fig:BelowLowerBound}
\end{figure}

In Fig.~\ref{fig:BelowLowerBound} we illustrate this generic feature
by a simple example. The black dashed
line depicts the lower bound
within $\phi^4$ theory ($\lambda_{2,\Lambda}=\lambda_{3,\Lambda}=0$),
whereas the red solid line shows Higgs masses for the initial data
$\lambda_{2,\Lambda}=-0.1$ and $\lambda_{3,\Lambda}=3$. This example
flow shows that the lower Higgs-mass bound can be significantly
relaxed if the restriction to bare potentials of $\phi^4$ type is
dropped. We emphasize that this restriction to renormalizable
operators is meaningless for the bare action as Wilsonian
renormalizability arguments do not apply to the bare field theory
action that might be generated from an unknown underlying UV complete
theory. 

Similarly to the $\mathbb{Z}_2$ invariant model, this phenomenon of a
relaxed bound as a consequence of a modified bare theory can be
understood by the RG flow itself \cite{Gies:2013fua}.  First note that
the parameters for the generalized bare potential are chosen in such
a way that the potential is initially in the SYM regime; this is also
true for the lower bound within $\phi^4$ theory. In the present case
of the generalized bare potential, the negative quartic coupling
$\lambda_2$ flows quickly to positive values whereas $\lambda_3$
becomes small as expected in the vicinity of the Gau\ss ian fixed point.
Therefore, the system essentially flows back into the class of
$\phi^4$-type potentials.  In other words, the system defined for a
fixed cutoff $\Lambda$ with $\lambda_{2,\Lambda}<0$ and
$\lambda_{3,\Lambda}>0$ can be mapped to a system with
$\lambda_{2,\Lambda}>0$ and $\lambda_{3,\Lambda}\approx 0$ for an
effectively smaller cutoff $\tilde\Lambda<\Lambda$.  Roughly speaking
some RG time is required to flow from the beyond-$\phi^4$-type
potentials back to the class of standard $\phi^4$-type
potentials. Thereby, the red dashed line can be interpreted as a
horizontally shifted variant of the Higgs mass curve derived from
$\phi^4$ potentials for effectively larger cutoffs. 

We emphasize that the effective potential is stable at all scales with
one well defined minimum for the present choice of
parameters.\footnote{As we compute the effective potential by means of
  a polynomial expansion about the minimum, these polynomials can
  seemingly develop new minima or instabilities at extremely large
  field values as we vary the truncation order $\Np$. In
  \cite{Gies:2013fua}, we have therefore carefully estimated the
  convergence radius of this expansion. For all examples presented
  here, the effective potentials does not show any instability or
  second minimum within this radius of convergence where the
  polynomial expansion can be trusted. This is confirmed by numerical
  integrations of the flow equation for the full effective potential
  as performed in \cite{Gneiting:2005} using pseudo-spectral methods
  (Chebyshev expansion).} Finally, we would like to point out that the
nonperturbatively computed effective potential for a finite cutoff is
of course regularization scheme dependent much in the same way as the
fermion determinant presented above. Choosing different regulator
shape functions would also lead to (typically slightly) different
Higgs mass curves in Figs.~\ref{fig:IRWindowHiggsMassWithinPhi4} -
\ref{fig:BelowLowerBound}. We expect that this scheme change could be
compensated for by a corresponding change of the bare action. Hence,
it suffices to vary only the bare action for a fixed regulator in
order to illustrate our main points.

\section{Conclusions}
\label{sec:conc}

\subsection{Summary}
We have analyzed a chiral Yukawa model featuring the interactions of a
scalar SU(2) Higgs field with a chiral top-bottom quark sector similar
to the Higgs sector of the standard model. We have critically
re-examined conventional perturbative arguments that relate a lower
bound for the Higgs mass with the stability of the effective
potential. Based on exact results for the regularized fermion
determinant, we have shown that the interacting part of the fermion
determinant contributes strictly positively to the effective scalar
potential for any finite field value -- as long as the UV cutoff
$\Lambda$ is kept finite. We have shown that this result holds for a
variety of regularization schemes including the sharp momentum cutoff
as well as (gauge-invariant) $\zeta$-function/proper-time
regularization schemes. 

Furthermore, we have shown that the conventional perturbative
conclusion of a vacuum in-/metastability of the effective potential
due to top-fluctuations can be rediscovered if the cutoff is forced to
approach infinity together with standard \textit{ad hoc} recipes to
project onto the finite parts. For the example of the sharp cutoff, we
have shown explicitly that this corresponds to an illegitimate order
of limits, as the resulting instability occurs at scalar field values
where the supposedly small expansion parameter of the
$\Lambda\to\infty$ limit is actually of order one. A similar failure
occurs for dimensional regularization where the standard procedures of
projecting onto the finite parts violate the positivity properties of
the interacting part of the effective potential. Our findings
corroborate earlier results from nonperturbative lattice simulations
\cite{Holland:2003jr,Holland:2004sd}, but in addition allow for a
large separation of the UV cutoff from the Fermi scale and an analytic
control of the corresponding limits.

Because of the presumable triviality of the present model as well as
the Higgs sector of the full standard model, the cutoff most likely
cannot be removed from the theory -- at least not within a
straightforward manner. The cutoff as well as a corresponding
regularization scheme should rather be viewed as part of the
definition of our particle physics models that parametrize the
embedding of this field-theory description into a possibly UV complete
theory. Still, as long as the cutoff is large compared to the Fermi
scale, Wilsonian renormalization arguments guarantee that the
low-energy observables are largely insensitive to the details of this
embedding. We have demonstrated that a counter-example to this generic
rule is given by bounds on the mass of the Higgs boson.

In this work, we have not performed an exhaustive analysis of
different bare actions or potentials, but simply focused on a
constructive example that leads to Higgs boson masses below the
conventional lower bound. Most importantly, this example exhibits no
vacuum in-/metastability.

This together with our basic line of argument involving exact results
for the fermion determinant demonstrate that there is no reason for
concern arising from top-quark fluctuations as far as false vacuum
decay in our universe is concerned, despite the comparatively light
value of the measured Higgs mass. This does not mean that there might
be no reason for concern at all. For instance, if the bare scalar
potential itself features an instability induced by the underlying UV
complete theory, our standard model could still live in an un- or
metastable vacuum. Our arguments only exclude instabilities caused by
the fluctuations of the fermionic matter fields within the standard
model.

\subsection{Vacuum stability vs. consistency bounds}

Our results suggest a revision of the standard picture of Higgs mass
bounds as a function of the UV cutoff. Depending on the implicit
assumptions made to derive mass bounds, this revision might be more or
less significant:

From our results on fermion determinants, it is clear that the
conventional interpretation that top-quark fluctuations induce a
vacuum instability is not tenable, but a result of taking an
inconsistent $\Lambda\to\infty$ limit. Still, the top-quark
fluctuations play, of course, a decisive role for the value of the
Higgs mass. In order to reconcile these observations, we propose a
UV-to-IR viewpoint: the Higgs mass bounds should be understood as a
mapping from initial conditions set at the UV cutoff given in terms of
a microscopic bare action $S_\Lambda$ onto all IR values accessible
by the RG flow of the system, $\mH=\mH[\Lambda;S_\Lambda]$. In this
manner, Higgs mass bounds arise from consistency conditions imposed on
the bare action. For instance, in order to start from a well-defined
(Euclidean) partition function, the action needs to be bounded from
below. 

The conventional vacuum stability bounds then are approximately
equivalent to such a consistency bound arising within a restricted
class of bare actions, e.g., bare potentials of $\phi^4$ type; here,
the bare $\phi^4$ coupling is required to be positive for consistency
of the generating functional. However, as the bare action is not at
our disposal but generally a result of the underlying UV embedding,
there is no reason to make such restrictive assumptions. Already for
slightly more general bare actions, we have been able to show that the
conventional lower mass bounds can be substantially relaxed in the
present chiral Yukawa model. The reason is that a more general bare
action can modify the RG flow near and below the cutoff. As a result,
the consistency bound lies below the vacuum stability bound. In
particular, we have given an explicit example with a Higgs boson mass
below the ``stability bound'' but an in fact stable effective
potential on all scales; our results are in agreement with lattice 
simulations \cite{Hegde:2013mks} and extend them to a much wider range
of scales.

Determining the consistency bound remains an open problem, the
solution of which requires further assumptions. One natural but not
necessary assumption could be that the effective action should feature
a unique minimum on all scales. The consistency bound would then arise
from a complicated extremization problem in the space of all
consistent bare actions subject to the unique-minimum constraint (to
be satisfied on all scales). Even in this case, it seems unclear
whether the bound remains finite. Therefore, it appears reasonable to
add another physical assumption: since the bare action is expected to
be provided by an underlying (UV complete) theory at scale $\Lambda$,
it is natural to assume (in the absence of any concrete knowledge
about the underlying theory) that the couplings of all possible
operators are of order $\mathcal{O}(1)$ if measured in terms of
$\Lambda$. For future studies, it is one of the most pressing
questions to quantitatively estimate the resulting consistency bound
under such a set of assumptions.

Of course, it appears equally legitimate to give up the criterion of a
unique minimum at all scales, but instead allow for further minima in
the bare action. If the resulting IR effective action turns out to
have one unique minimum again (to be identified with the electroweak
minimum), such bare actions can lead to a further relaxation of the
consistency bound described above. In the general case, it should be
possible to construct bare actions with multiple local minima such
that the full effective action has a global minimum different from the
local electroweak minimum. Since such bare actions are less
constrained than those of the preceding scenarios, we expect the
resulting lower Higgs mass consistency bounds to be even more relaxed
to smaller values. Again, a quantitative estimate of such consistency
bounds including meta-stable scenarios remains an urgent question.

Comparing the conventional stability bounds with the present
consistency bounds, the overall picture seems to be qualitatively
similar. The primary main difference is of quantitative nature, since
the unique-minimum consistency bound lies below the stability
bound. Wilsonian RG arguments however suggest that this difference
could become small for large UV cutoffs, as is also reflected by the
example of Fig.~\ref{fig:BelowLowerBound}. Nevertheless, the size of this
quantitative difference substantially depends on the assumptions
imposed on the size of the couplings in the bare action. Also, the
consistency bound is necessarily regularization scheme dependent. As
the regularization actually should model the details of the embedding
into the underlying UV completion, this dependence has a physical
meaning. Even larger differences are expected between the conventional
meta-stability bound and the consistency bound including meta-stable
scenarios. The reason is that the meta-stable features are expected to
arise from the bare action and thus are largely unknown. The size of
the meta-stable region and corresponding life-time estimates will be
even more subject to assumptions on the bare action.

\subsection{Outlook}

Independently of whether the measured value of the Higgs boson mass
eventually turns out to lie slightly above or below the conventional
lower bound, it is remarkable that the Higgs and top mass parameters
appear to lie close to a region in the IR parameter space that can be
connected to a bare UV effective potential that could exhibit almost
vanishing scalar self-interactions. In this sense, a precise
measurement of these mass parameters is relevant beyond the pure goal
of precision data. These measurements can impose requirements that any
UV embedding has to satisfy. The viewpoint of consistency bounds
presented above provides a means to quantify these
requirements. Therefore, a comprehensive quantitative exploration of
these bounds appears most pressing.

One scenario appears particularly interesting: if the Higgs mass
eventually turns out to be exactly compatible with a UV flat potential
(apart from a possible mass term), a corresponding embedding would
have to explain this rather particular feature. It is interesting to
note that such scenarios exist even within purely quantum field theory
approaches as, for instance, in models with asymptotically safe
gravity \cite{Shaposhnikov:2009pv,Bezrukov:2012sa}.

Recently, an asymptotically safe/free scenario in a gauged chiral
Yukawa model has been identified \cite{Gies:2013pma}, the UV limit of
which corresponds to a flat scalar potential also allowing for
comparatively light Higgs masses in the IR. In order to explore this
option of a UV complete limit, we perform an RG fixed point search
also within this model along the lines of \cite{Gies:2009sv} in
App. \ref{appB}. However, in this ungauged model, we find no reliable
indication for the existence of a non-Gau\ss{}ian fixed
point. Still, our present findings should serve as a strong
motivation to further search for asymptotically free gauged chiral
models with viable low-energy properties.

\section*{Acknowledgments}

We thank Gerald Dunne, Tobias Hellwig, J\"org J\"ackel, Stefan
Lippoldt, Axel Maas, Jan Pawlowski, Tilman Plehn, Michael Scherer,
Andreas Wipf, Christof Wetterich, and Luca Zambelli for interesting
discussions. We acknowledge support by the DFG under grants
No. GRK1523/2, and Gi 328/5-2 (Heisenberg program).

\appendix

\section{Threshold functions}
\label{appA}

The threshold functions $l,m$ in the flow equations for the various
couplings parametrize the decoupling of massive modes. As a function
of their mass-type arguments, they approach zero in the limit of large
masses and tend to a constant value in the zero-mass limit. The
explicit form of these threshold functions depends on the regulator,
characterizing the details of the momentum-shell integration.  In this
work, we use the linear regulator \cite{Litim:2001up} which allows to
work out the threshold functions analytically. For the bosonic modes,
this regulator is given by
\begin{align*}
R_k(p)= Z_{\phi} p^2 \, r(p^2/k^2) = Z_{\phi}(k^2-p^2)\theta(k^2-p^2).
\end{align*}
The corresponding chirally symmetric fermionic regulator
$R_k(p)=Z_{\psi} \fss{p} r_{\text{F}}(p^2/k^2)$ is chosen such that
$p^2(1+r)=p^2(1+r_{\text{F}})^2$.  For reasons of completeness, we
list the threshold functions appearing in the main text for the linear
regulator:
\begin{align*}
 l_n^d(\omega) &= \frac{4(\delta_{n,0}+n)}{d}\frac{1-\frac{\eta_{\phi}}{d+2}}{(1+\omega)^{n+1}},
 \\
 l_{0\, \text{L}}^{(\text{F})d}(\omega) &= \frac{4(\delta_{n,0}+n)}{d} \frac{1-\frac{\eta_{L}}{d+1}}{(1+\omega)^{n+1}},
 \\
 l_{0\, \text{R}_{\text{t/b}}}^{(\text{F})d}(\omega) &= \frac{4(\delta_{n,0}+n)}{d} \frac{1-\frac{\eta_{R}^{t/b}}{d+1}}{(1+\omega)^{n+1}},
 \\
 l_{n_1,n_2}^{(\text{FB})d}(\omega_1,\omega_2;\eta_{\psi},\eta_{\phi}) &= \frac{2}{d}\frac{1}{(1+\omega_1)^{n_1}}\frac{1}{(1+\omega_2)^{n_2}} \\
 &\hspace{-0.45cm}\times \left[ \frac{n_1 \left(1-\frac{\eta_{\psi}}{d+1}\right)}{1+\omega_1} 
         + \frac{n_2 \left(1-\frac{\eta_{\phi}}{d+2}\right)}{1+\omega_2} \right],
 \\
 m_{n_1,n_2}^d(\omega_1,\omega_2) &= \frac{1}{(1+\omega_1)^{n_1}(1+\omega_2)^{n_2}},
 \\
 m_2^{(\text{F})d}(\omega) &= \frac{1}{(1+\omega)^4},
\end{align*}
\begin{align*}
 m_4^{(\text{F})d}(\omega;\eta_{\psi}) &= \frac{1}{(1+\omega)^4} + \frac{1-\eta_{\psi}}{d-2}\frac{1}{(1+\omega)^3} \\ 
 &\quad - \left( \frac{1-\eta_{\psi}}{2d-4} + \frac{1}{4} \right) \frac{1}{(1+\omega)^2},
 \\
 m_{n_1,n_2}^{(\text{FB})d}(\omega_1,\omega_2;\eta_{\psi},\eta_{\phi}) &= \frac{1-\frac{\eta_{\phi}}{d+1}}{(1+\omega_1)^{n_1}(1+\omega_2)^{n_2}}.
\end{align*}
These threshold functions agree with those given in  \cite{Hofling:2002hj}.

\section{Mean-field effective potential}
\label{sec:MF}

Let us generalize our results for the fermionic determinant obtained
for a sharp cutoff in the main text to a wider class of regulator
shape functions in this section. This corresponds to a mean-field
analysis of the effective scalar potential for general momentum-space
regularization schemes. More technically speaking we integrate
Eq.~(\ref{flowpot}) for fixed Yukawa couplings, $h_a \to
h_{a,\Lambda}$ ($a=\{\text{t,b}\}$), as well as fixed wave function
renormalizations, $Z_i \to 1$, where the subscript $i$ labels the
different fields, from $k=\Lambda$ to $k=0$. In addition we also
neglect the bosonic contributions on the right-hand side of
Eq.~(\ref{flowpot}). The resulting mean-field potential reads:
\begin{align}
 U_k^{\text{MF}}(\rho) = U_{\Lambda}(\rho) + d_{\text{W}} \int_p \sum_{a=\{\text{t},\text{b}\}} \ln{ \frac{p^2(1+r_{\text{F},\Lambda})^2 + h_{a,\Lambda}^2\rho}{p^2(1+r_{\text{F},k})^2 + h_{a,\Lambda}^2\rho} },
 \label{eq:mean-field}
\end{align}
where the regulator shape
function $r_{\text{F}}$ depends on the momentum as well as on the RG
scale, $r_{\text{F},k}=r_{\text{F}}(p^2/k^2)$. In order to
provide for a regularization, $r_\text{F}(x)$ should vanish for
large argument and diverge sufficiently fast to positive infinity
for $x\to 0$. Apart from analyticity for all finite $x>0$, no
further requirements on $r_{\text{F}}$ are needed; however, for an
interpretation of a physical regularization, we assume
$r_{\text{F}}$ to be positive for finite $x$.  The second
derivative of the mean-field effective potential with respect to
$\rho$ encodes the fermionic contributions to the interacting part of
the effective potential:
\begin{align*}
 &{U_{k=0}^{\text{MF}}}''(\rho) = U_{\Lambda}''(\rho) + \frac{d_{\text{W}}}{8\pi^2} \sum_{a=\{\text{t},\text{b}\}} h_{a,\Lambda}^4 \times \\
 &\times\!\! 
 \int_0^{\infty}\!\!\!\!\!\! dp\, p^3 \frac{[p^2(1+r_{\text{F},\Lambda})^2-p^2][p^2(1+r_{\text{F},\Lambda})^2 + p^2 + 2h_{a,\Lambda}^2\rho]}{[p^2(1+r_{\text{F},\Lambda})^2+h_{a,\Lambda}^2\rho]^2 [p^2+h_{a,\Lambda}^2\rho]^2}.
\end{align*}
The integrand is strictly positive for
$r_{\text{F}}>0$ and so is the integral.  This corroborates the conclusion
in the main text that the interaction part of the fermion determinant
is strictly positive for a wide class of regulators.

Furthermore it is instructive to calculate the Higgs boson mass in
this simple approximation. For the linear regulator the mean-field
effective potential for $k=0$ reads,
\begin{align*}
 U^{\text{MF}} &= U_{\Lambda} + \frac{d_{\text{W}}}{32\pi^2} \times \\
 &\quad \times\hspace{-0.3cm} \sum_{a=\{\text{t},\text{b}\}} \left[ -h_{a,\Lambda}^2\rho\Lambda^2 + (h_{a,\Lambda}^2\rho)^2 \ln{\left(1+\frac{\Lambda^2}{h_{a,\Lambda}^2\rho}\right)} \right],
\end{align*}
where the decomposition into a negative mass-like term $\sim\rho$ and a positive interaction part is manifest. The Higgs boson
mass is now given as the second derivative of the potential at the
nonvanishing minimum $v=246$GeV (${U^{\text{MF}}}'(v)=0$):
\begin{align*}
 \mH^2 &= 2\rho \frac{\partial^2 U^{\text{MF}}}{\partial\rho^2} \Bigg|_{\rho=v^2/2} \\
 &= \frac{\mtop^4}{4\pi^2v^2} \left[ 2\ln{\left(1+\frac{\Lambda^2}{\mtop^2}\right)} - \frac{3\Lambda^4+2\Lambda^2\mtop^2}{(\Lambda^2+\mtop^2)^2} \right] \\
 &\quad + \frac{\mbot^4}{4\pi^2v^2} \left[ 2\ln{\left(1+\frac{\Lambda^2}{\mbot^2}\right)} - \frac{3\Lambda^4+2\Lambda^2\mbot^2}{(\Lambda^2+\mbot^2)^2} \right] \\
 &\quad + v^2U_{\Lambda}''\Big(\frac{v^2}{2}\Big),
\end{align*}
where we have used $d_{\text{W}}=2$ for 2-component Weyl fermions.
Here it is obvious that the restricted class of quartic bare
potentials ($U_{\Lambda}''=\lL\geq 0$) gives rise to a lower bound for
the Higgs mass for $\lL=0$, nonetheless no instability is induced by
the fermionic fluctuations.

\section{Cut-Off mechanism for the Goldstone modes}
\label{CutOffGoldstones}

In the main text, we have amended the chiral Yukawa model with a
dynamical mechanism that effectively removes the Goldstone bosons from
the low-energy spectrum by providing them with a mass of the order of
the gauge-boson mass in the full standard model. Conceptually, this
mechanism can be viewed as an IR deformation of our model and is
neither universal nor unique. Here, we explore the sensitivity of our
results for the Higgs boson mass on the details of this mechanism.

\begin{figure}
 \centering
 \includegraphics[width=8.5cm]{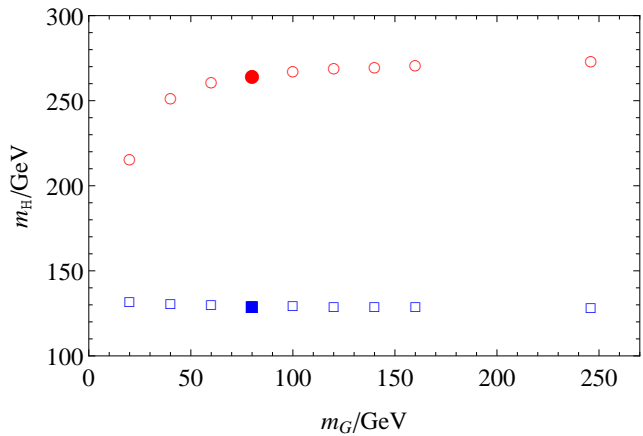}
 \caption{Higgs mass $\mH$ as a function of the modeled mass of the
   would-be Goldstone bosons for a weakly coupled ($\lambda_{2,\Lambda}=0$, blue
   squares) as well as for a strongly coupled ($\lambda_{2,\Lambda}=10$,
   red circles) scalar sector. The filled characters mark the Higgs
   masses computed for the value $m_G=80$GeV which we have used in the
   main text to derive our quantitative results.}
 \label{fig:HiggsVSGoldstone}
\end{figure}

Within our deformation, let us vary the parameter $g$, which controls
the size of the effective mass for the Goldstone bosons proportional
to the vev, $m_{\text{G}}^2 = (1/2) g v_k^2$.  Figure
\ref{fig:HiggsVSGoldstone} shows the Higgs mass depending on the
effective Goldstone mass at $\Lambda=10^7$GeV.  For a weakly coupled
scalar sector, $\bar\lambda_{\Lambda}=0$ (blue squares), the impact of
$g$ on $\mH$ is insignificant.  Even if the Goldstone mass changes by
an order of magnitude the influence on the Higgs mass is less than 2 GeV. 
The situation is different for a strong scalar coupling, e.g.,
$\bar\lambda_{\Lambda}=10$ (red circles).  For large effective
Goldstone masses ($m_{\text{G}}>\mH$), we observe again only small
deviations from the case $m_{\text{G}}=80$GeV. This is because the
Goldstone modes decouple roughly at the same scale as the top and the
Higgs.  However, if we choose smaller values of $m_{\text{G}}$ the
deviations become larger, e.g., we observe a deviation of $20\%$ for
$m_{\text{G}}=20$GeV compared to the case of $m_{\text{G}}=80$GeV.
Now, the Goldstone modes contribute over a wider range of scales to
the flow equations than the Higgs or the top quark. This results in a
smaller Higgs mass.  For vanishing $m_{\text{G}}$ we observe a
log-like running of $\lambda_2$ (which is presumably an artifact of
our definition of this coupling in terms of a Cartesian field
decomposition). As a result, $\mH$ approaches smaller and smaller
values in the limit $g\to 0$.

Of course, beyond the deformation of the model chosen in this work,
other methods are conceivable to effectively remove the massless
Goldstone bosons to mimic the physics of the fully gauged standard
model.  One option is to choose an arbitrary scale $k_{\text{nG}}$
within the SSB regime of the flow, below which all Goldstone boson
contributions to the flow equations for RG scales smaller than
$k_{\text{nG}}$. This corresponds to switching on a source at
$k_{\text{nG}}$ which gives the Goldstone modes an infinitely large
mass and leads to an \textit{ad hoc} decoupling.  Another possibility is
to introduce a source term $J^a\phi^a$ on the level of the
Lagrangian. This would leave the flow equations unchanged because only
the Hessian of the effective average action $\Gamma_k^{(2)}$
contributes to the Wetterich equation. Still, the shape of the scalar
potential would be changed by this explicit symmetry breaking term. As
a result, source-dependent mass terms for all scalar degrees of
freedom are induced, and also the vev as well as the couplings become
source-dependent, $\kappa=\kappa(J)$, $\lambda_n=\lambda_n(J)$.  In
any case, each of these different IR deformations of the model leads
to similar results. They parametrize a decoupling of the Goldstone
modes removing the contamination of the IR flow by particle degrees of
freedom which would not be present in the standard model.
%
%
%

\section{Impact of the top quark mass on Higgs mass values}

The top mass plays an important role in the study of the lower Higgs
mass bound within $\phi^4$ theory. Within the perturbative line of
reasoning, a change of the top mass by 1 GeV goes along with a change
of the Higgs mass bound by approximately 2 GeV in standard model
calculations. As the Higgs mass is near (or presumably below) the
conventional lower bound, an accurate determination of the mass of the
top quark in the appropriate renormalization scheme
\cite{Alekhin:2012py} is crucial for a discussion of the consequences
of this ``near criticality'' \cite{Buttazzo:2013uya}.

In this appendix, we study the values for the Higgs boson mass for a
given initial bare potential of $\phi^4$ as a function of the top
quark mass in order to analyze the top quark mass dependence in the
present model. Within our truncation of the effective action, our
top-mass IR parameter automatically coincides with the pole mass, the
latter being the phenomenologically relevant
quantity. (Straightforwardly computable) differences between these
mass parameters only arise at next-to-next-to-leading order in the
derivative expansion. 

\begin{figure}
 \centering
 \includegraphics[width=8.5cm]{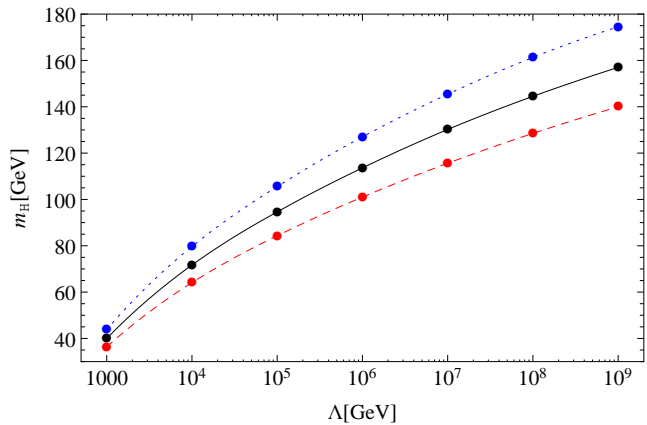}
 \caption{Higgs mass $\mH$ as a function of the cutoff $\Lambda$ for
   different values of $\mtop$. The black solid line is the lower
   bound for $\mtop=173$GeV using $\lambda_{2,\Lambda}=0$ for a
   $\phi^4$-type bare potential. In addition lower Higgs mass bounds
   for $\mtop=163$GeV (red dashed lower curve) and $\mtop=183$GeV
   (blue dotted upper line) are plotted.}
 \label{fig:VaryTopMass}
\end{figure}

\begin{table}
 \begin{tabular}{|c|c|c|}
  \hline
  $\mtop$[GeV] & $\mH$[GeV] $(\lambda_{2,\Lambda}=0)$ & $\mH$[GeV] $(\lambda_{2,\Lambda}=10)$
  \\ \hline\hline
  163 & $115.6$ & $264.2$ \\ \hline
  165 & $118.5$ & $264.6$ \\ \hline
  167 & $121.4$ & $264.8$ \\ \hline
  169 & $124.2$ & $265.0$ \\ \hline
  171 & $127.1$ & $265.2$ \\ \hline
  \textbf{173} & $\textbf{130.1}$ & $\textbf{265.4}$ \\ \hline
  175 & $133.1$ & $265.7$ \\ \hline
  177 & $136.1$ & $265.9$ \\ \hline
  179 & $139.3$ & $266.2$ \\ \hline
  181 & $142.2$ & $266.4$ \\ \hline
  183 & $145.4$ & $266.7$ \\ \hline
 \end{tabular}
 \caption{Higgs masses for the lower bound ($\lambda_{2,\Lambda}=0$)
   and for a strongly interacting scalar sector
   ($\lambda_{2,\Lambda}=10$) for a cutoff of $\Lambda=10^7$GeV as a
   function of the top quark mass. Bold numbers indicate the standard
   top mass of 173GeV.}
 \label{tab:Higgs_Top}
\end{table}
For the class of $\phi^4$-type bare potentials, table
\ref{tab:Higgs_Top} summarizes our results for the Higgs mass for the
lower mass bound ($\lambda_{2,\Lambda}=0$) as well as for a strongly
coupled scalar sector ($\lambda_{2,\Lambda}=10$) as a function of the
top quark mass for $\Lambda = 10^7$GeV. For the weakly coupled scalar
sector, $\lambda_{2,\Lambda}=0$, the Higgs mass is generated
essentially through top quark fluctuations.  Therefore, the resulting
Higgs mass values are most sensitive to the precise mass of the top
quark. In the present model, the Higgs mass is shifted by approximately
3 GeV for a change in the top mass by 2 GeV using a cutoff of
$\Lambda=10^7$GeV.  With increasing cutoff the deviation of the Higgs
mass is larger. This phenomenon is illustrated in
Fig. \ref{fig:VaryTopMass} where the spread of the lower bound for
larger cutoff scales is shown.

By contrast, we find only a slight impact of the top mass on the Higgs
mass for the case of a strongly coupled scalar sector. This
illustrates that the Higgs mass is rather dominated by scalar
fluctuations in this regime.

\section{Fixed point structure}
\label{appB}

As the flow equation provides us with information about the system
beyond perturbative limitations, we can explore the properties of the
model also at stronger coupling. In particular, it is worthwhile to
search for possible RG fixed points at finite values of the couplings
(non-Gau\ss{}ian fixed points), as these offer the chance to evade the
triviality problem. Pure field-theory UV completions are possible
within the asymptotic safety scenario, where the UV limit
$\Lambda\to\infty$ can be taken safely by means of a UV stable RG
fixed point \cite{Weinberg:1976xy}, as is even explored for quantum
gravity \cite{Reuter:1996cp}, see
\cite{Percacci:2007sz,Braun:2010tt} for reviews and
examples. Provided such a fixed point exists and has suitable
properties the theory remains interacting (non-trivial) in the
long-range limit and has predictive power.

The search for such fixed points in Yukawa systems has recently been
revived with systematic studies in the framework of the functional
renormalization group. A general mechanism for inducing asymptotic
safety in such systems has been identified in \cite{Gies:2009hq},
relying on a dynamical balance between boson and fermion
fluctuations. Chiral Yukawa models have successively been studied in
this context \cite{Gies:2009sv} providing hints for the possible
existence of such fixed points, but also indicating that fully gauged
models may be required for a stable asymptotic safety scenario of
standard-model like theories. Gauged models indeed appear to offer
different routes to UV complete theories either in the weak-coupling
asymptotically-free limit \cite{Gies:2013pma}, at fully interacting
fixed points (including the loss of asymptotic freedom)
\cite{Litim:2014uca}, or even in combination with quantum gravity
\cite{Zanusso:2009bs,Dona:2013qba}. In fact, asymptotic safety has
become a viable concept of consistently quantizing gravity in the
recent years. Also nonlinear chiral models have been explored along
this direction \cite{Bazzocchi:2011vr}.

Here, we concentrate again on the pure chiral Yukawa sector,
performing an analysis much in the spirit of \cite{Gies:2009sv},
paying particular attention to the additional bottom quark degree of freedom.

\subsection{Symmetric Regime}
Whereas the mechanism identified in \cite{Gies:2009hq} operates in the
regime of spontaneous symmetry breaking, it is instructive to start
the fixed point search in the symmetric regime, where the flow
equations are less complex.
%
The flow equations for the Yukawa couplings using the linear regulator
read in this regime:
\begin{align*}
 \pt \htop^2 &= \htop^2 \Bigg[ \eta_{\phi}+2\eta_{\text{t}} \\
 &\qquad\quad - \hbot^2\frac{16v_d}{d}\left( \frac{1-\frac{\eta_{\text{t}}}{d+1}}{1+\lambda_1}+\frac{1-\frac{\eta_{\phi}}{d+2}}{(1+\lambda_1)^2} \right) \Bigg],
 \\
 \pt \hbot^2 &= \hbot^2 \Bigg[ \eta_{\phi}+2\eta_{\text{b}} \\
 &\qquad\quad - \htop^2\frac{16v_d}{d} \left( \frac{1-\frac{\eta_{\text{b}}}{d+1}}{1+\lambda_1}+\frac{1-\frac{\eta_\phi}{d+2}}{(1+\lambda_1)^2} \right) \Bigg],
\end{align*}
where $\lambda_1\geq0$ corresponds to the mass-like term in the
effective potential in the SYM regime.  Let us start with resolving
the fixed point conditions $\pt \htop^2=0$ and $\pt \hbot^2=0$ in leading
order in the derivative expansion: for $\eta_i=0$, both conditions can
be satisfied for either $\hbot=0$ and $\htop$ arbitrary or vice
versa. Without loss of generality, let us assume that $\hbot=0$, leaving
us with a free parameter $\htop^\ast$, labeling a potential line of
fixed points. Now it is easy also to solve the fixed-point condition
for the effective potential, $\pt u = 0$, cf. \Eqref{flowpot}, with an
arbitrarily chosen value for $\htop^\ast$ at least within a
polynomial expansion of $u(\rho)$. We do not pursue this any further,
since this line of fixed points does not exist beyond leading order.

At next-to leading order it is necessary to include the equations for
the anomalous dimensions, which read for the linear regulator in the
SYM regime
\begin{align*}
 \eta_{\phi} &= \frac{8v_d}{d}\big[ \htop^2(4-\eta_{\text{t}})+\hbot^2(4-\eta_{\text{b}}) \big], \\
 \eta_{\text{t}} &= \frac{4v_d}{d}\frac{3\htop^2+\hbot^2}{(1+\lambda_1)^2}\Big(1-\frac{\eta_{\phi}}{d+1}\Big), \\
 \eta_{\text{b}} &= \frac{4v_d}{d}\frac{\htop^2+3\hbot^2}{(1+\lambda_1)^2}\Big(1-\frac{\eta_{\phi}}{d+1}\Big).
\end{align*}
It is useful to study a linear
combination of the two fixed point conditions, $0 = \frac{1}{\htop^2}\pt
\htop^2 - \frac{1}{\hbot^2}\pt \hbot^2$, assuming that $\htop\neq 0 \neq \hbot$,
\begin{align}
 0 &= \frac{1}{\htop^2}\pt \htop^2 - \frac{1}{\hbot^2}\pt \hbot^2 \notag \\
 &= 2(\eta_{\text{t}}+\eta_{\text{b}}) - \frac{16v_d}{d} \left[ \frac{1-\frac{\eta_{\text{b}}}{d+1}}{1+\lambda_1}+\frac{1-\frac{\eta_\phi}{d+2}}{(1+\lambda_1)^2} \right] (\htop^2-\hbot^2) \notag \\
 &= -\frac{8v_d}{d} \left[ 2\frac{1-\frac{\eta_{\text{t}}}{d+1}}{1+\lambda_1} + \frac{1-\frac{d\eta_{\phi}}{(d+2)(d+1)}}{(1+\lambda_1)^2} \right] (\htop^2-\hbot^2).
 \label{symfixp}
\end{align}
In order to justify the use of the derivative expansion of the
effective action, we demand for the anomalous dimensions $\eta_i$ to
remain sufficiently small also at a possible fixed point, $\eta_i^\ast
\lesssim 1$. As a consequence, the term in the square bracket of
\Eqref{symfixp} is positive. This condition can only be solved by
$\htop^2=\hbot^2$. Therefore the system of algebraic fixed-point
equations reduces to
\begin{equation}
\begin{split}
 0 &= \eta_{\phi}+2\eta_{\text{t}} - \frac{16v_d}{d}\htop^2\Big[\frac{1-\frac{\eta_{\text{t}}}{d+1}}{1+\lambda_1}+\frac{1-\frac{\eta_{\phi}}{d+2}}{(1+\lambda_1)^2}\Big],
 \\
 \eta_{\text{t}} &= \frac{16v_d}{d}\htop^2\frac{1-\frac{\eta_{\phi}}{d+1}}{(1+\lambda_1)^2},
 \\
 \eta_{\phi} &= \frac{16v_d}{d} \htop^2(4-\eta_{\text{t}}).
 \end{split}
\label{symfixp2}
\end{equation}
The constraints on the anomalous dimensions $\eta_i<1$ imply that also
$\eta_i> 0$ holds, as the negative terms linear in $\eta_i$ on
the right-hand sides of \Eqref{symfixp2} remain subdominant compared
with the positive terms. Expressing the constraints
$0<\eta_{\phi}<1$ and $0<\eta_{t}<1$ through a constraint on the top
Yukawa coupling via the last two lines of \Eqref{symfixp2} leads to
a constraint on $\htop^2$: $0\leq \htop^2 \leq
4\pi^2[5(1+\lambda_1)^2-\sqrt{5}\sqrt{(1+\lambda_1)^4-(1+\lambda_1)^2}]$
for $d=4$. Within this constraint, a solution for the fixed point
equation (first line of \Eqref{symfixp2}) does not exist independently
of any permissible value of $\lambda_1$. Hence, we do not find a
nontrivial fixed point within our present truncation in the symmetric
regime in four spacetime dimensions.
\subsection{SSB regime}
Because of the couplings of the fluctuating fields to the
condensate the structure of the flow equations in the SSB
regime is much richer than in the symmetric case. As a starting
point we analyze the beta functions to leading order in the derivative
expansion and expand the potential to the lowest nontrivial order in
the field invariant, $u=\frac{\lambda_2}{2}(\rho-\kappa)^2$. Therefore,
we have to solve the following nonlinear system of equations:
\begin{alignat*}{3}
 \pt \kappa &= && \beta_{\kappa}(\kappa^*,\lambda_2^*,\htop^*,\hbot^*) && = 0,\\
 \pt \lambda_2 &= && \beta_{\lambda_2}(\kappa^*,\lambda_2^*,\htop^*,\hbot^*) && = 0,\\
 \pt \htop &= && \beta_{\htop}(\kappa^*,\lambda_2^*,\htop^*,\hbot^*) && = 0,\\
 \pt \hbot &= && \beta_{\hbot}(\kappa^*,\lambda_2^*,\htop^*,\hbot^*) && = 0,
\end{alignat*}
where the flow equations can be read off from Eqs.~\eqref{flowpot} and
\eqref{flowYukawa}. For the linear regulator, this system can
be solved analytically resulting in two inequivalent fixed
points for admissible physical parameters ($\kappa>0$, $\htop^2\geq
0$, $\hbot^2\geq 0$, $\lambda_2>0$). The values of the fixed point
couplings are listed in Tab.~\ref{tab:fixpoint}.  First of all note
that there are two additional fixed points by exchanging the numerical
values for $\htop^*$ and $\hbot^*$ due to the fact that the flow
equations are symmetric under an exchange of the Yukawa couplings. Of
course, these additional fixed points are physically equivalent to
those given in Tab.~\ref{tab:fixpoint}, corresponding to a mere
renaming of the couplings.

\begin{table}[h]
 \begin{tabular}{|c|c|c|c|}
  \hline
  $\kappa^*$ & $\lambda_2^*$ & ${\htop^*}^2$ & ${\hbot^*}^2$ \\ \hline\hline
  0.006749086 & 14.6058 & 6942.84 & 0 \\ \hline
  0.000934843 & 270.652 & 661.201 & 0 \\ \hline
 \end{tabular}
 \caption{Fixed point values for the various parameters in the SSB regime to leading order in the derivative expansion.}
 \label{tab:fixpoint}
\end{table}

Furthermore, the flow equations of the Yukawa couplings are
proportional to the Yukawa couplings themselves, $\pt h_a^2 \propto
h_a^2$, which leads to a decoupling of the bottom quark at all
scales. Therefore the system reduces to that studied in
\cite{Gies:2009sv}.

In order to check if these fixed points persist beyond the
leading order, we have to include the anomalous dimensions of the
fields. As long as these quantities, which measure the influence of
higher derivative terms in our truncation, remain small, the
derivative expansion appears legitimate. For a first impression
of the size of the anomalous dimensions, we insert the leading order
fixed point values listed in Tab.~\ref{tab:fixpoint} into the
right-hand sides of Eqs.~(\ref{eq:anomalous})-(\ref{eq:anomalous2}).
Similar to the results of \cite{Gies:2009sv} the anomalous dimensions
of the fields at the fixed points are large, especially
$\eta_{\text{t}}$ is much larger than one, ($\eta_{\text{t}}\simeq 22$
for the first fixed point and $\eta_{\text{t}}\simeq 4$ for the second
one). This casts serious doubts on the existence of these fixed
points in the full theory. In fact, it has been shown numerically in
\cite{Gies:2009sv} that the fixed points do not persist as soon as
back-reactions of the anomalous dimensions are included. This can be
traced back to large contributions of massless modes, such as the
Goldstone and bottom quark modes near the would-be fixed point.

To summarize, we find no indications that the present chiral
model in its pure form supports physically acceptable fixed points
within the validity domain of the derivative expansion of the
effective action.

\end{document}